\documentclass[twocolumn]{aastex62} 
\usepackage{graphicx}
\usepackage{txfonts}
\usepackage{threeparttable}

\received{xxx}
\revised{xxx}
\accepted{}
\submitjournal{ApJ}
\shorttitle{A hard look at NGC 5347}
\shortauthors{Kammoun et al.}

\newcommand{  \Hbeta    }{\ifmmode {\rm H}\beta \else H$\beta$\fi}
\newcommand{  \Halpha   }{\ifmmode {\rm H}\alpha \else H$\alpha$\fi}
\newcommand{  \caii     }{\ifmmode {\rm Ca}\,\textsc{ii}   \else Ca\,\textsc{ii}\fi}
\newcommand{\Cahk}{\ifmmode \left[{\rm Ca H+K}\,\textsc{ii}\right\,\lambda3935,3968 \else Ca H+K$\,\lambda3935,3968$\fi}
\newcommand{\Mgb}{\ifmmode \left{\rm Mg}\,\textsc{i}\right\,\lambda5175 \else Mg\,{\sc i}\,$\lambda5175$\fi}
\begin{document} 

\title{A hard look at NGC 5347: revealing a nearby Compton-thick AGN}
\correspondingauthor{E. Kammoun}
\email{ekammoun@umich.edu}

\author[0000-0002-0273-218X]{E. S. Kammoun}
\affiliation{Department of Astronomy, University of Michigan, 1085 South University Avenue, Ann Arbor, MI 48109-1107, USA}

\author{J. M. Miller}
\affiliation{Department of Astronomy, University of Michigan, 1085 South University Avenue, Ann Arbor, MI 48109-1107, USA}

\author{A. Zoghbi}
\affiliation{Department of Astronomy, University of Michigan, 1085 South University Avenue, Ann Arbor, MI 48109-1107, USA}

\author{K. Oh}
\altaffiliation{JSPS fellow}
\affiliation{Department of Astronomy, Kyoto University, Oiwake-cho, Sakyo-ku, Kyoto 606-8502, Japan}

\author{M. Koss}
\affiliation{Eureka Scientific, 2452 Delmer Street Suite 100, Oakland, CA 94602-3017, USA}

\author{R. F. Mushotzky}
\affiliation{Department of Astronomy and Joint Space-Science Institute, University of Maryland, College Park, MD 20742, USA}

\author{L. W. Brenneman}
\affiliation{Harvard-Smithsonian Center for Astrophysics, 60 Garden St., Cambridge, MA 02138, USA}

\author{W. N. Brandt}
\affiliation{Department of Astronomy and Astrophysics, 525 Davey Lab, The Pennsylvania State University, University Park, PA 16802, USA}
\affiliation{Institute for Gravitation and the Cosmos, The Pennsylvania State University, University Park, PA 16802, USA}
\affiliation{Department of Physics, 104 Davey Lab, The Pennsylvania State University, University Park, PA 16802, USA}

\author{D. Proga}
\affiliation{Department of Physics \& Astronomy, University of Nevada Las Vegas, Las Vegas, NV 89154, USA}

\author{A. M. Lohfink}
\affiliation{Montana State University, P.O. Box 173840, Bozeman, MT 59717-3840, USA}

\author{J. S. Kaastra}
\affiliation{SRON Netherlands Institute for Space Research, Sorbonnelaan 2, 3584 CA Utrecht, the Netherlands}
\affiliation{Leiden Observatory, Leiden University, PO Box 9513, 2300 RA Leiden, the Netherlands}

\author{D. Barret}
\affiliation{IRAP, Universit\'{e} de Toulouse, CNRS, UPS, CNES, 9, Avenue du Colonel Roche, BP 44346, 31028 Toulouse Cedex 4, France}

\author{E. Behar}
\affiliation{Department of Physics, Technion, 32000, Haifa, Israel}

\author{D. Stern}
\affiliation{Jet Propulsion Laboratory, California Institute of Technology, 4800 Oak Grove Drive, MS 169-221, Pasadena, CA 91109, USA}

\begin{abstract}

Current measurements show that the observed fraction of Compton-thick (CT) AGN is smaller than the expected values needed to explain the cosmic X-ray background. Prior fits to the X-ray spectrum of the nearby Seyfert-2 galaxy NGC 5347 ($z=0.00792,\, D =35.5 \rm ~Mpc $) have alternately suggested a CT and Compton-thin source. Combining archival data from \textit{Suzaku}, \textit{Chandra}, and - most importantly - new data from \textit{NuSTAR}, and using three distinct families of models, we show that NGC 5347 is an obscured CTAGN ($N_{\rm H} > 2.23\times 10^{24}~\rm cm^{-2}$). Its 2-30~keV spectrum is dominated by reprocessed emission from distant material, characterized by a strong Fe K$\alpha$ line and a Compton hump. We found a large equivalent width of the Fe K$\alpha$ line ($\rm EW = 2.3 \pm 0.3$~keV) and a high intrinsic-to-observed flux ratio ($\sim 100$). All of these observations are typical for bona fide CTAGN. We estimate a bolometric luminosity of $L_{\rm bol} \simeq 0.014 \pm 0.005~L_{\rm Edd.}$. The \textit{Chandra} image of NGC 5347 reveals the presence of extended emission dominating the soft X-ray spectrum ($E < 2~\rm keV$), which coincides with the [\ion{O}{3}] emission detected in \textit{Hubble Space Telescope} images. Comparison to other CTAGN suggests that NGC 5347 is broadly consistent with the average properties of this source class. We simulated \textit{XRISM} and \textit{Athena}/X-IFU spectra of the source, showing the potential of these future missions in identifying CTAGN in the soft X-rays.

\end{abstract}

\keywords{galaxies: active --- galaxies: individual (NGC 5347) --- galaxies: Seyfert --- X-rays: general}

%
\section{Introduction}
\label{sec:intro}

It is well accepted that active galactic nuclei (AGN) are powered by the accretion of matter onto a supermassive black hole (SMBH) through a geometrically thin, optically thick disk \citep[e.g.,][]{Shak73}. The ``unified model'' of AGN \citep{Antonucci93, Netzer15} hypothesizes the presence of a dusty circumnuclear torus at the parsec scale, explaining the dichotomy between type-1 and type-2 AGN through different viewing angles. The actual morphology and composition of this material is an open question, although several works suggest a clumpy distribution of optically-thick clouds rather than a homogeneous structure \citep[e.g.,][]{Honig07, Risaliti07, Balokovic14, Marinucci16}.

A significant fraction ($\sim 10-25 \%$) of the AGN population, in the local Universe, is theoretically expected to be obscured by Compton-thick (CT) material (with an equivalent neutral hydrogen column density, $N_{\rm H} \gtrsim 1.5 \times 10^{24}\,\rm cm^{-2} $) in order to explain the observed peak of the cosmic X-ray background (CXB) in the 20-50~keV band \citep[see e.g.,][and references therein]{Ueda14}. However, the observed fraction of CTAGN is much smaller than these values. Only about $8\%$ of the AGN in the {\it Swift}/BAT 70-month catalog are found to be CT \citep[see][]{Ricci15}. The major difficulty in identifying CTAGN is due to the fact that the emission in the soft X-rays, ultraviolet (UV), and optical, which is directly produced by the AGN, is heavily attenuated due to obscuration. The only two spectral bands where the obscuring material is optically thin up to high column densities are the hard X-rays ($\gtrsim 15$~keV) and the mid-infrared ($5-50\,\rm \mu m$). Thanks to its unprecedented sensitivity covering the $3-79$~keV band, \textit{NuSTAR} is playing a key role in identifying the missing fraction of CTAGN and determining their properties \citep[e.g.,][]{Koss16, Marchesi18}.

In this paper, we present multi-epoch observations of the Seyfert-2 galaxy NGC 5347 ($z=0.00792,\, D =35.5 \rm ~Mpc $) using the \textit{Chandra X-ray Observatory}, \textit{Suzaku}, and \textit{NuSTAR}. This source is part of a \textit{NuSTAR} Legacy Survey (PI: J. M. Miller) aiming to study an optically-selected volume-limited sample of 22 Seyfert-2 galaxies that were identified in the CfA Redshift Survey \citep{Huchra83}. \cite{Risaliti99} classified this source as a CTAGN ($N_{\rm H} > 10^{24}\,\rm cm^{-2}$) on the basis of its large Fe K$\alpha$ equivalent width ($\rm EW > 1.9~\rm keV$) measured in an \textit{ASCA} spectrum. However, \cite{Lamassa11} classified the source as Compton thin, with $N_{\rm H} = 5.6_{-2.3}^{+3.2}\times 10^{23}~\rm cm^{-2}$, based on the \textit{Chandra} spectrum. We note that this source is a megamaser galaxy showing strong [\ion{O}{4}] emission, negligible star formation and no hint of silicate absorption at 10~$\rm \mu m$ \citep{Hernan15}. The paper is organized in the following way: in Section \ref{sec:sdss} we present the analysis of an optical spectrum of the source. The X-ray observations are presented in Section \ref{sec:reduction}. We discuss the X-ray spectral fitting within the context of various models in Section \ref{sec:spectral_fits}. Finally, in Section \ref{sec:discussion} we discuss the implications of our results and we present spectral simulations of this source for the future high-resolution observatories: {\it XRISM} and {\it Athena}. The following cosmological parameters are assumed: $\Omega_{\rm M}= 0.27$, $\Omega_{\Lambda} =0.73$ and $H_0 = 70\,\rm km s^{-1}\rm Mpc^{-1}$.

\section{Optical spectroscopy}
\label{sec:sdss}

\begin{figure}
\centering

\includegraphics[width = 1.\linewidth]{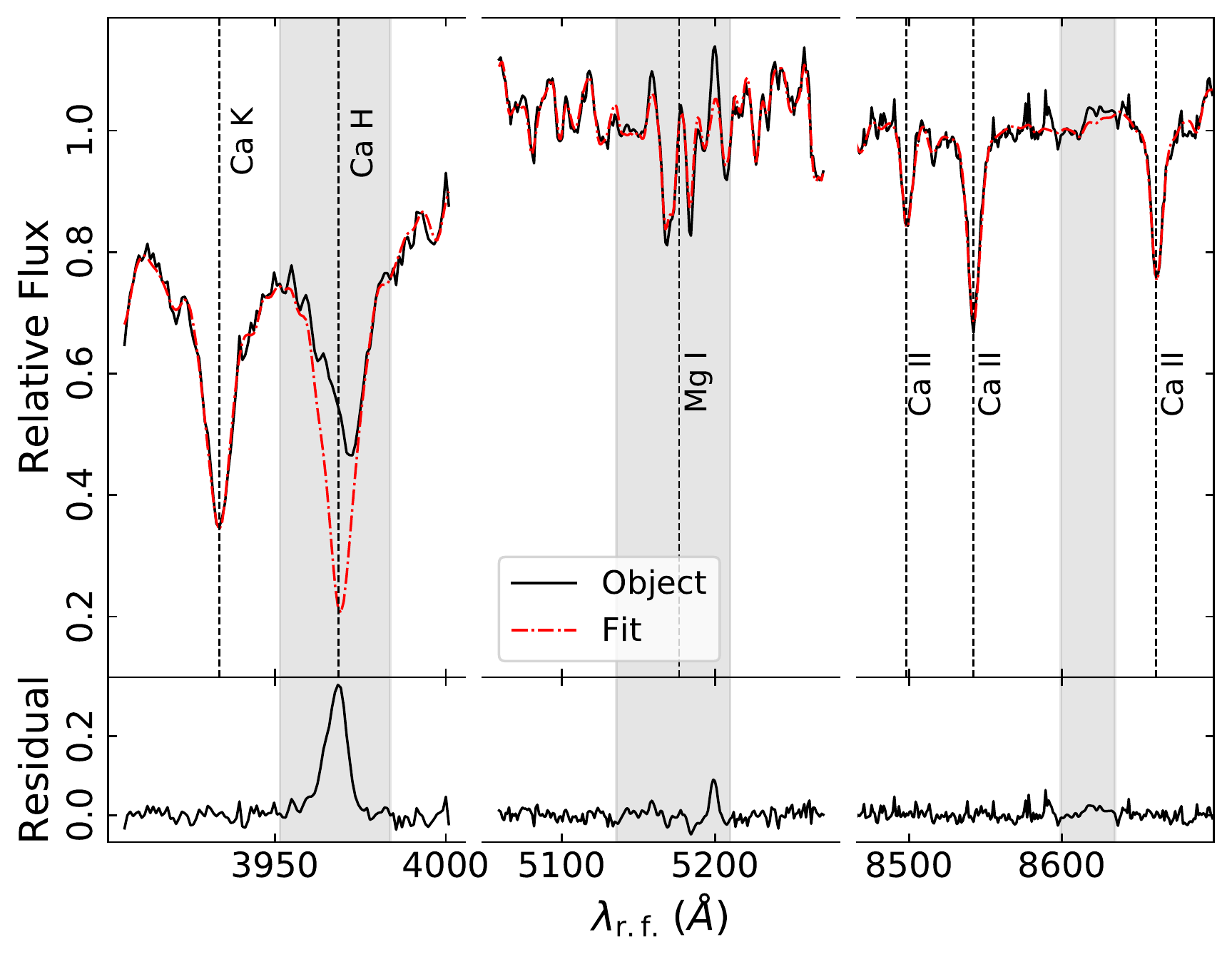}
\includegraphics[width = 1.\linewidth]{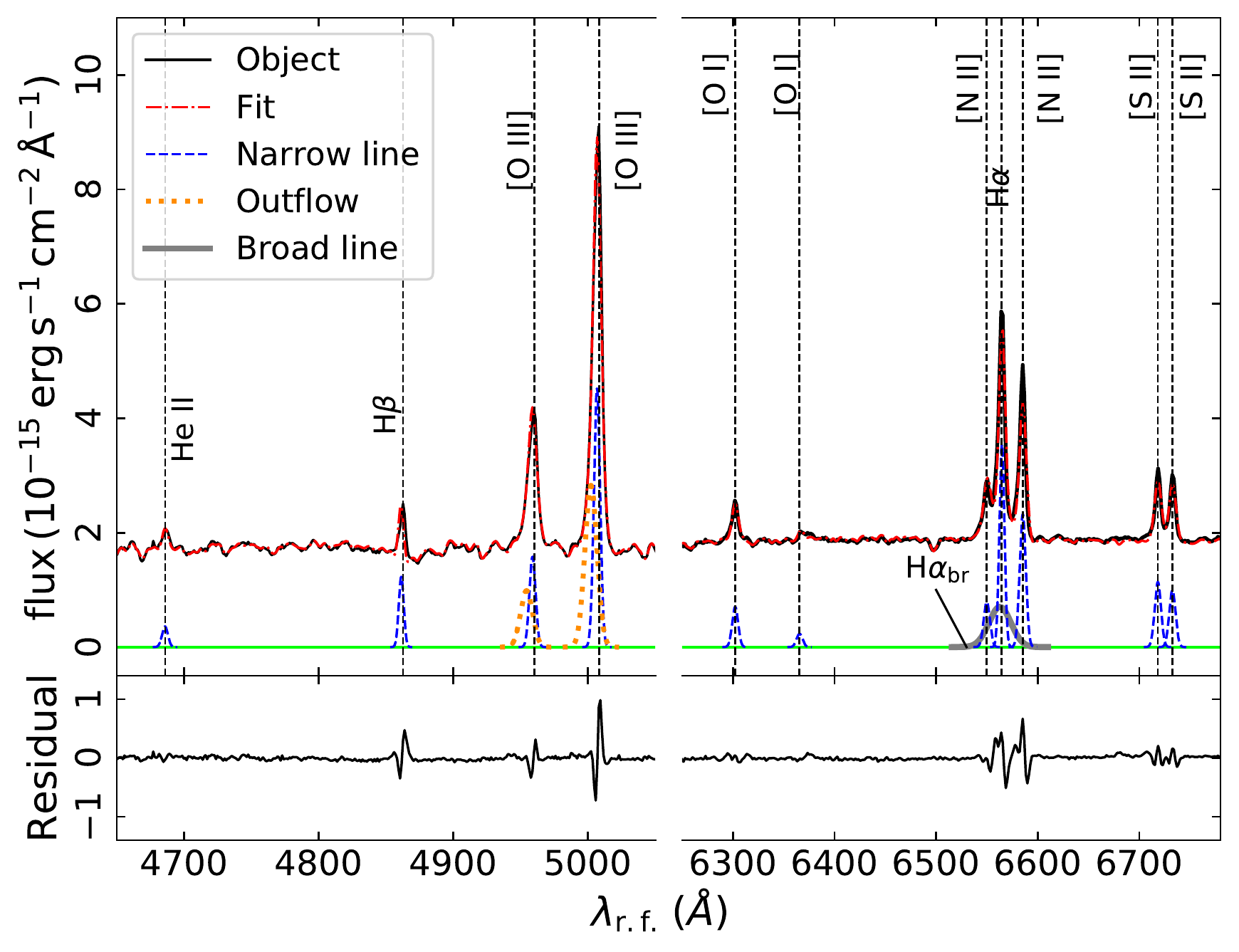}

\caption{SDSS spectrum of NGC~5347 (black solid lines) and the best-fit models (red dash-dotted lines). The upper panel shows the spectrum of NGC~5347 with the fits of the absorption lines in the \Cahk, \Mgb\, and the \caii\ triplet spectral spectral regions. The grey shaded areas represent the regions with emission lines which are excluded from the fit (e.g., Ca H$\lambda$3968.47 and [\ion{N}{1}]$\lambda$5200, [\ion{Fe}{2}]$\lambda$8619). The bottom panel shows the rest frame NGC~5347 spectra corrected for Galactic extinction in the \Halpha\ and the \Hbeta\ complexes.}
\label{fig:sdss}
\end{figure}

NGC~5347 was observed as part of the Sloan Digital Sky Survey (SDSS) on 2007-01-15 with a 3$''$ (513 pc) diameter fiber for a total exposure of 4203 seconds \citep{SDSSDR7}. The continuum and the absorption features were fit using the penalized PiXel Fitting software \citep[{\tt pPXF};][]{Cappellari04} to measure a central velocity dispersion for the galaxy. A stellar template library from VLT/Xshooter \citep{Chen14} was used with to fit the spectrum with optimal stellar templates following the general procedure in \cite{Koss17}. These templates have been observed at higher spectral resolution ($R=10,000$) than the AGN observations and are convolved in {\tt pPXF} to the spectral resolution of each observation before fitting. When fitting the stellar templates, all the prominent emission lines were masked (see upper panel of Fig.~\ref{fig:sdss}).

The main aim of studying the SDSS spectrum of this source is to determine the mass of the SMBH from the stellar velocity dispersion using a high-resolution spectrum. In fact, it has been shown that, for some cases, low-resolution spectra could lead to an overesitmate of the velocity dispersion \citep[e.g.,][]{Brightman18}. In the case of NGC 5347, we find a velocity dispersion of $89\pm3~\rm km~s^{-1}$ in the \Cahk\ and \Mgb\ regions (3830-5600~\AA) and $93\pm5~\rm km~s^{-1}$ for the \caii\ triplet spectral region (8350--8700\,\AA) for a weighted average of $90 \pm 3~\rm km~s^{-1}$. This measurement is consistent with the literature values \citep[$73\pm14 ~\rm km~s^{-1}$, $103 \pm 10 ~\rm km~s^{-1}$;][respectively]{Terlevich90,Nelson95} but  shows significantly less error. Using the \cite{Kormendy13} relation, this velocity dispersion implies a black hole mass of $\log\left( M_{\rm BH}/M_{\odot} \right) = 6.97 \pm 0.13$. This value is consistent with the measurement reported by \cite{Izumi16}, $\log \left( M_{\rm BH}/M_{\odot} \right) = 6.73 \pm 0.55$, from the velocity dispersion measurements above.

\begin{table}
\centering

\caption{Measured emission-line fluxes, Gaussian amplitude over noise, and EW for the major emission lines observed in the SDSS spectrum of NGC~5347. }

\begin{tabular}{llll}
\hline \hline
Line	&	Flux	&	A/N	&	EW \\ 
 & $(10^{17}~\rm erg~s^{-1}~cm^{-2})$ & & (\AA) \\ \hline

\ion{He}{2}$\lambda$4686 & $  217.59 \pm 2.06$ & 11.25 & 1.32 \\ 
\Hbeta & $650.68 \pm 5.77$ & 38.03 & 3.83 \\ 

[\ion{O}{3}]$\lambda$5007 & $2884.32 \pm 21.47$ & 139.60 & 16.91\\ 

[\ion{O}{1}]$\lambda$6300 &  $566.94 \pm 4.20 $ & 21.80 & 3.05 \\ 

[\ion{O}{1}]$\lambda$6363 & $190.80 \pm 1.40$ & 7.26 & 1.02 \\ 
\Halpha & $2528.77 \pm 16.49$ & 109.46 & 13.65 \\

[\ion{N}{2}]$\lambda$6583 & $1951.50 \pm 10.41$ & 71.81 & 10.68 \\ 

[\ion{S}{2}]$\lambda$6716 & $976.91 \pm 5.33$ & 35.24 & 5.32\\ 

[\ion{S}{2}]$\lambda$6730 & $853.55 \pm 4.63$ & 30.72 & 4.61 \\ \hline\hline
\end{tabular}
\label{table:optlines}
\end{table}

For emission line measurements, we follow the procedure used in the OSSY database \citep{Oh11} and its broad-line prescription \citep{Oh15}. We perform stellar templates \citep{Bruzual03, Sanchez-blazquez06} and emission-line fitting in a rest-frame ranging from 3780~\AA\ to 7250~\AA\ (see bottom panel of Fig.~\ref{fig:sdss}). We correct the narrow line ratios ($H\alpha/H\beta$) assuming an intrinsic ratio of $R = 3.1$ and the \cite{Cardelli89} reddening curve. The measured Balmer decrement H$\beta$/H$\alpha=3.9$ suggests a low level of extinction of the narrow line region (NLR) and is consistent with the Balmer decrement for most optically selected AGN from the SDSS \citep[see Fig. 12 in][]{Koss17}. We measure the AGN emission-line diagnostics and confirm that NGC 5347 is consistent with a Seyfert galaxy using the [\ion{O}{3}]/\Hbeta\ vs.~[\ion{N}{2}]/\Halpha, \ion{S}{2}/\Halpha, and [\ion{O}{1}]/\Halpha\ diagnostics \citep[e.g.,][]{Veilleux87, Kewley06}. The observed characteristics of the major emission lines are reported in Table~\ref{table:optlines}.

The [\ion{O}{3}] line shows a blue wing consistent with an NLR outflow consitent with the outflows found in the HST data \citep{Schmitt03} and the extended emission observed with \textit{Chandra} (see next section for more details). We note that past observations had found no broad emission lines in the optical spectrum of the source \citep{degrijp92}. However, we found some evidence of residuals in the \Halpha-[\ion{N}{2}] complex that may be consistent with a weak broad line around \Halpha\ [$\rm EW(H\alpha_{\rm br})  = 10.38~ \AA$]. High-velocity wings associated with the outflow may be present in both the \Halpha\ and [\ion{N}{2}] line profiles and thus could be misinterpreted as an underlying broad \Halpha\ component associated with the broad line region (BLR). We note that most studies reported much more significant equivalent width of broad lines  \citep[e.g., $\rm \langle EW \rangle \sim 70\,\AA$;][]{Shen11}. \cite{Oh15} reported substantial number of unidentified weak broad-line in type 1 AGN in the local Universe ($z<0.2$) whose $\rm EW(H\alpha_{\rm br}) $ is peaking at $\sim 30~\AA$ \citep[see figure 9 in][]{Oh15}. To further understand the NLR outflows in this system and better detect or constrain possible weak broad lines, further observations would be required, such as high spatial resolution integral field NIR spectroscopy or spectropolarimetry.

\section{X-ray Observations and data reduction}
\label{sec:reduction}
NGC 5347 was observed by the \textit{Chandra X-ray Observatory} on 2004-06-05 (PI: N. Levenson; ObsID 4867), by \textit{Suzaku} on 2008-06-10 (ObsID 703011010), and by \textit{NuSTAR} (ObsID 60001163002) on 2015-01-16 . The log of the observations is presented in Table~\ref{table:log}. Here, we summarize our data reduction procedures.

\begin{table}
\centering

\caption{Net exposure time, average net count rate and the ratio of the source to total counts, in the observed 3--10\,keV band.}

\begin{tabular}{lccc}
\hline \hline
Instrument	&	Net exposure	&	Count rate	&	Source/total	\\
	&	(ks)	&	(count ks$^{-1}$)	&		\\ \hline
\textit{Chandra} (ACIS)	&	36.9	&	$4.44 \pm 0.36$	&	97.0~\%	\\
\textit{Suzaku} (XIS-FI)	&	42.0 	&	$3.17 \pm 0.44$	&	33.0~\%\\
\textit{NuSTAR} (FPMA)	&	46.5	&	$4.53 \pm 0.34$	&	84.8~\%	\\
\textit{NuSTAR} (FPMB)	&	46.6	&	$3.04 \pm 0.30$	&	76.1~\%	\\ \hline\hline

\end{tabular}
\label{table:log}
\end{table}
\subsection{\textit{ Suzaku} observations}

The XIS \citep{XIS} spectra from {\it Suzaku} \citep{Suzaku07} were reduced following standard procedures using {\tt HEASOFT}. The initial reduction was done with {\tt aepipeline}, using the CALDB calibration release v20160616. Source spectra were extracted using {\tt xselect} from circular regions 3$\arcmin$ in radius centered on the source. Background spectra were extracted from a source-free region of the same size, away from the calibration source. The response files were generated using {\tt xisresp}. We do not consider the spectrum from XIS1, owing to its poor relative calibration. Spectra from XIS0 and XIS3 were checked for consistency and then combined to form the front-illuminated spectra.

\subsection{\textit{ Chandra} observations}
\label{sec:chandra}

The \textit{Chandra} \citep{CXO00} data were reduced using CIAO version 4.9 and the latest associated calibration files. The source was observed close to the optical axis and nominal aimpoint on the backside-illuminated ACIS-S3 chip, meaning that the full spatial resolution of \textit{Chandra} can be exploited.

\begin{figure}
\centering
\includegraphics[width = 0.45\textwidth]{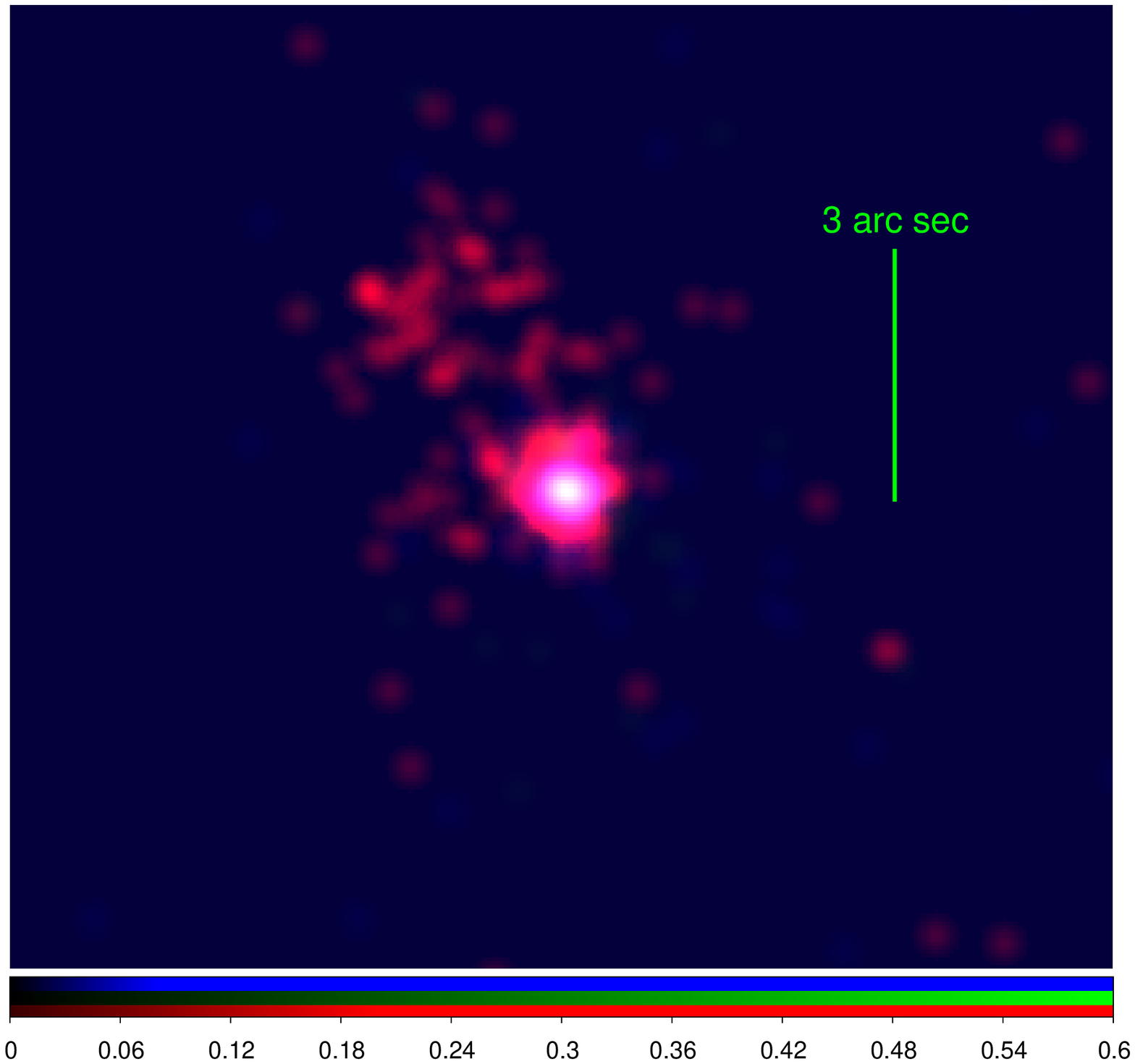}\\
\includegraphics[width = 0.45\textwidth]{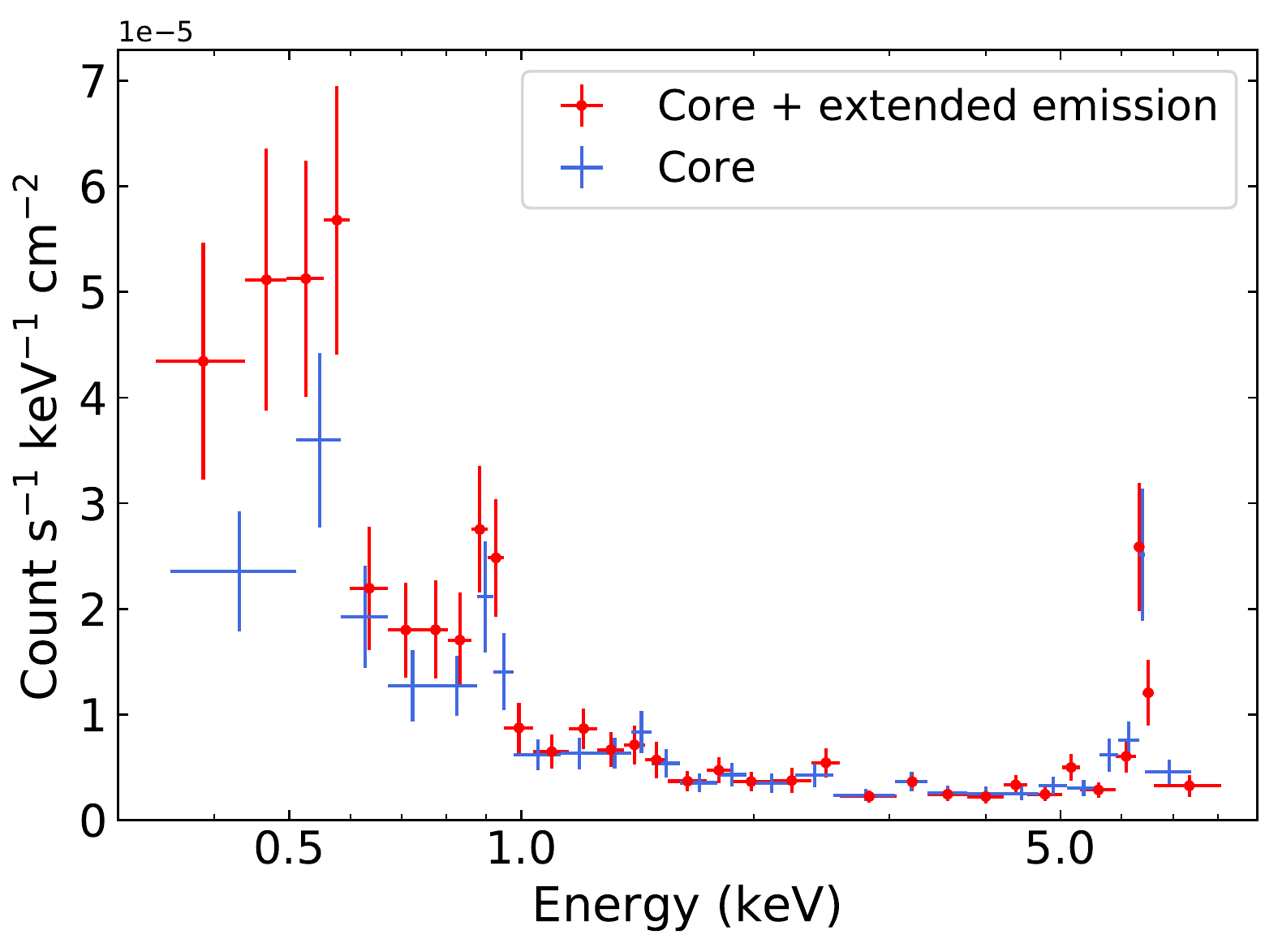}
\caption{Top panel: \textit{Chandra} image of NGCC~5347 showing the 0.3-1.2~keV band (red), 1.2-2.4~keV band (green) and 2.4-8~keV band (blue). The image shows a clear spatial extension of the emission in the 0.3-1.2~keV band. \textit{Chandra} spectra extracted from the core of the source (blue points) and from the core and the extended emission (red points; see Sect.~\ref{sec:chandra}). We note that the O K emission (below $\sim 0.6$~keV) is stronger in the spectrum when the extended emission was included.}
\label{fig:Chandraspectra}
\end{figure}

Prior work noted that NGC 5347 is slightly extended in the \textit{Chandra} image. The diffuse emission region closely coincides with [\ion{O}{3}] emission detected in \textit{Hubble Space Telescope (HST)} images, potentially indicating the direction of an ionization cone or the NLR \citep[][]{Schmitt03, Levenson06}. Using sub-pixel event reprocessing and energy filtering, we are able to confirm that the extended X-ray emission is strongest in the soft band ($E \leq 1.5$~keV), and potentially strongest of all in the O K band (below $\sim 0.6$~keV), as shown in Fig.~\ref{fig:Chandraspectra}. The flux ratios of the spectrum including the extended region over the spectrum the core region are $1.35 \pm 0.10$ and $1.09 \pm 0.11$, below and above 1.5~keV, respectively.

Source and background spectral files and response files were all created using the CIAO tool {\tt specextract}. We first extracted source counts from a circular region with a radius of 1.5\arcsec\ (256~pc) centered on the known source coordinates, ignoring the extended emission. We then extracted the source and diffuse emission jointly using a 5.2\arcsec\ (891 pc) circle, centered on the extended image. The resultant data were grouped to require at least 10 counts per spectral bin.  The neutral Fe K line is clearly evident in the spectrum, and is several times stronger than the local continuum (see Fig.~\ref{fig:Chandraspectra}); this indicates that the central engine is highly obscured and signals that the source is CT. It is notable that the spectrum including the extended X-ray and [\ion{O}{3}] emission region has more flux in the O K range, consistent with neutral oxygen at 0.525~keV, or a low-ionization charge state. The flux of the Fe line in the spectrum, including the extended emission, is consistent with the one from the core, indicating that the Fe line is emitted within an inner region of 1.5\arcsec. For consistency, we use the spectrum that includes the extended emission in the rest of the analysis.

\subsection{{\it NuSTAR} observations}
The {\it NuSTAR} \citep{Harrison13} data were reduced following the standard pipeline in the {\it NuSTAR} Data Analysis Software ({\tt NUSTARDAS}\,v1.8.0), and using the latest calibration files. We cleaned the unfiltered event files with the standard depth correction. We reprocessed the data using the ${\tt saamode = optimized}$ and ${\tt tentacle = yes}$ criteria for a more conservative treatment of the high background levels in the proximity of the South Atlantic Anomaly. We extracted the source and background spectra from circular regions of radii 45\arcsec\ and 100\arcsec, respectively, for both focal plane modules (FPMA and FPMB) using the {\tt HEASOFT} task {\tt Nuproduct}, and requiring a minimum S/N of 3 per energy bin. The spectra extracted from both modules are consistent with each other. The data from FPMA and FPMB are analyzed jointly in this work, but they are not combined together.

\section{X-ray Spectral analysis}
\label{sec:spectral_fits}

\begin{figure*}
\centering

\includegraphics[width = 0.49\linewidth]{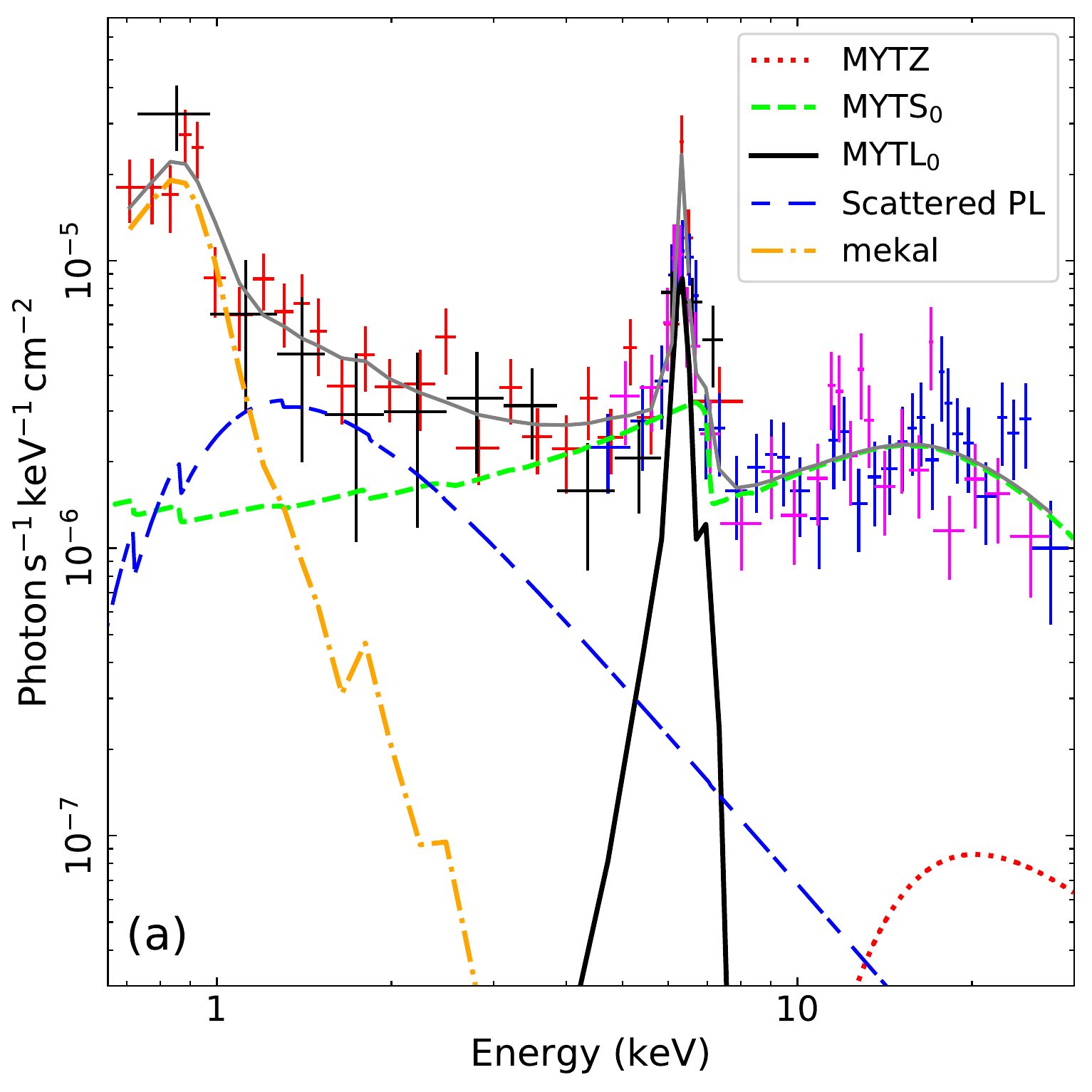}
\includegraphics[width = 0.49\linewidth]{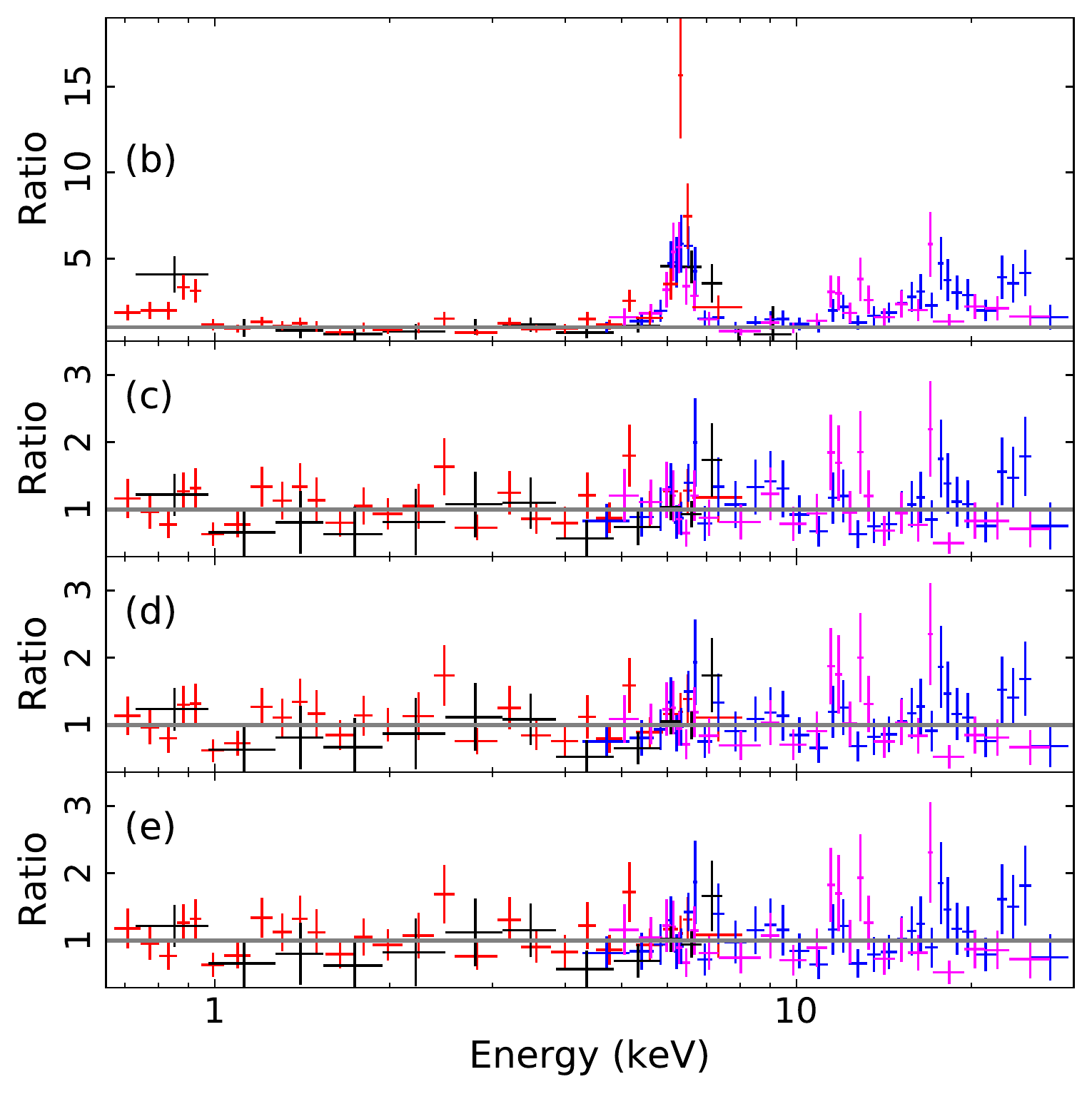}\\
\includegraphics[width = 0.6\linewidth]{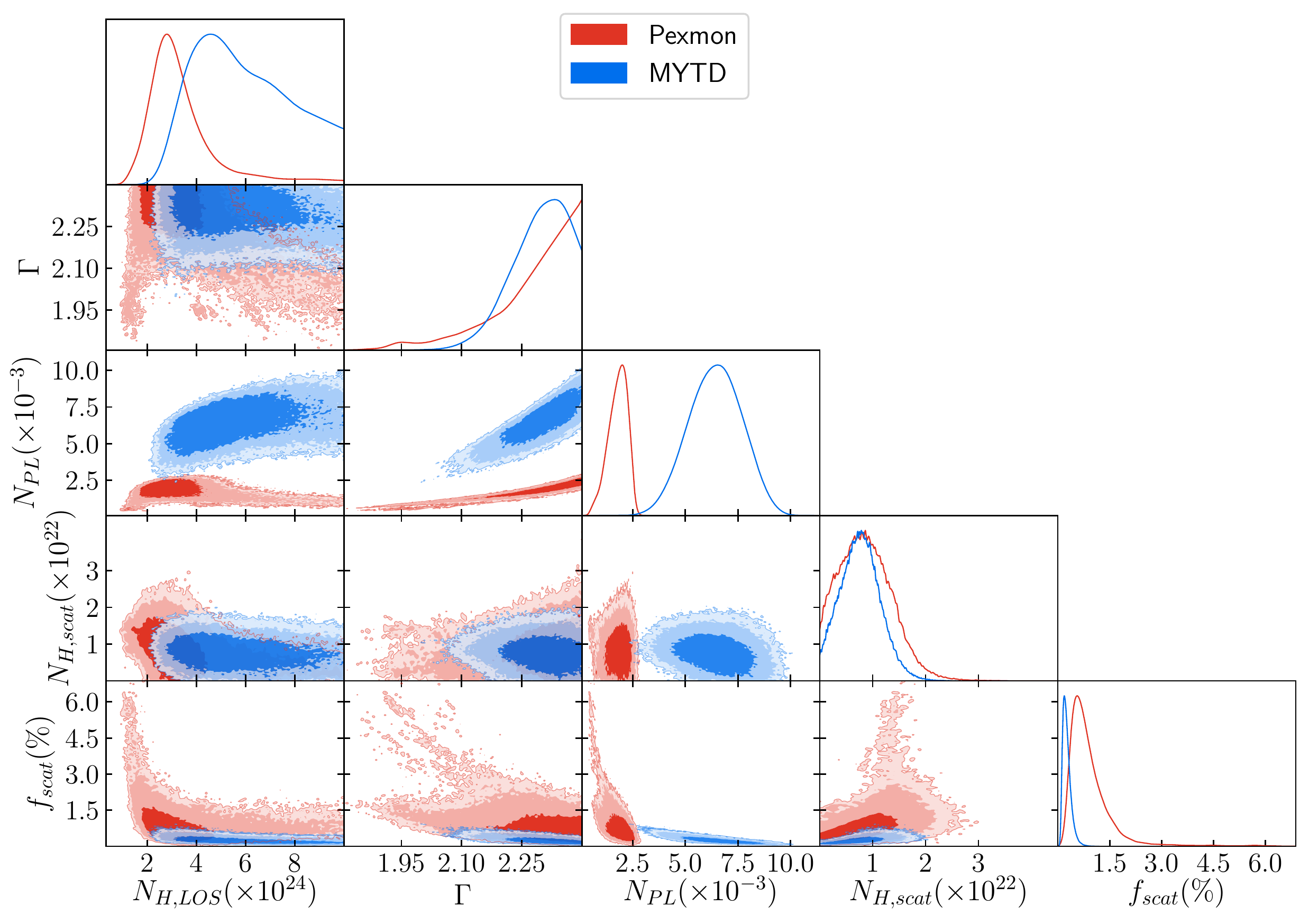}

\caption{ Upper panel: \textit{Suzaku} (black), \textit{Chandra} (red) and \textit{NuSTAR} FPMA/B (blue/magenta) spectra of NGC 5347. The grey line in panel (a) represents the best-fit MYTD model, in addition to the various components in the MYTD model. Panel (b) shows the ratio (data/model) by fitting the 1.5-5 keV band using a simple power law with $\Gamma = 0.85$. Panels (c), (d), and (e) show the residuals obtained by fitting Pexmon, MYTC and MYTD models, respectively. Lower panel: the confidence contours (showing the 1-$\sigma$, 2-$\sigma$ and 3-$\sigma$ levels) obtained from the MCMC analysis for the relevant parameters of the Pexmon (red) and MYTD (blue) models. The last panel in each row corresponds to the normalized 1-D probability density function of the corresponding parameter.}
\label{fig:spectra_model}
\end{figure*}

Throughout this work, spectral fitting was performed using XSPEC\,v12.10e \citep{Arnaud96}. Due to the energy limitation of some of the employed spectral models in this work and the data quality, we considered the \textit{Chandra} and the \textit{Suzaku} spectra in the observed 0.6--8 keV and 0.7-- 7.5 keV bands, respectively. The \textit{NuSTAR} spectra are background dominated below 4~keV and above 30~keV. For that reason, we fit the \textit{NuSTAR} data in the  4--29~keV band. Given the consistency between the various instruments (Fig.~\ref{fig:spectra_model}a), we fixed the cross calibration between them to unity. Throughout this work, we apply the Cash statistic \citep[$C$-stat;][]{Cash79}. Unless stated otherwise, uncertainties on the parameters are listed at the 1$\sigma$ confidence level ($\Delta C= 1$). These uncertainties, for all the models, are calculated from a Markov chain Monte Carlo (MCMC) analysis, starting from the best-fitting model that we obtained. We used the Goodman-Weare algorithm \citep{Goodman10} with a chain of $10^6$ elements discarding the first 30\% of elements as part of the ‘burn-in’ period. We note that due to the non-Gaussianity in the distribution of some parameters, the best-fit value found using XSPEC does not match with the mean value found in the chains, however they are consistent within $2\sigma$.

First, we fit the spectra in the $1-5$ keV band, using an absorbed power-law model accounting for the Galactic absorption in the line of sight (LOS) of the source \citep[$ N_{\rm H, Gal} =  1.52 \times 10^{20}\, \rm cm^{-2}$;][]{Kalberla05}. The model fits the data well ($C/{\rm dof} = 19.3/20$) with a hard photon index $\Gamma = 0.81 \pm 0.14$. Such a hard spectrum indicates the presence of strong absorption. The extrapolation of the best-fit model to the $0.6-30$ keV band reveals the presence of an excess in the soft X-rays, a strong excess in the Fe-line band, and a broader excess in the $10-30$~keV band, as shown in Fig.~\ref{fig:spectra_model}b. The former component is mainly due to thermal diffuse emission, while the latter two components give strong indication that the hard X-rays in this source are dominated by reprocessed emission. In the following, we present a detailed analysis of the X-ray spectra (in the $0.6-30$~keV range) by considering different models in order to describe the reprocessed emission in this source.

\subsection{Pexmon}
\label{sec:Pexmon}

We initially fit the spectra using the neutral reflection model Pexmon \citep{Nandra07}. The model can be written (in the XSPEC terminology) as follows:

\begin{eqnarray}
\begin{array}{lc}
{\tt model_{Pexmon} } & =  {\tt phabs[1] * ( zphabs[2]*cutoffpl[3]   }
\end{array}\nonumber\\
\begin{array}{lll}
  {\tt + zphabs[4]*constant[5]*cutoffpl[6]  } & 
\end{array}\nonumber\\
\begin{array}{lc}
{\tt + pexmon[7] + mekal[8]}  ).&\nonumber 
\end{array}
\label{eq:pexmon_model}
\end{eqnarray}

\noindent In this model, the {\tt phabs[1]} component represents the Galactic absorption, {\tt cutoffpl[3]} represents the primary emission of the source assumed to be a power-law with a high-energy cutoff (fixed to 500~keV), which is intrinsically absorbed by CT material ({\tt zphabs[2]}). A fraction (${\tt constant[5] *  cutoffpl[6]}$, where $0 \leq {\tt constant[5]} \leq 1$) of the primary emission could be scattered into our LOS, by optically thin ionized gas in the polar regions, before being possibly absorbed as well ({\tt zphabs[4]}). All the parameters of {\tt cutoffpl[6]} are tied to the ones of {\tt cutoffpl[3]}. The photon index, cutoff energy and normalization of {\tt pexmon[7]}, which describes the reprocess emission, are tied to the same parameters of {\tt cutoffpl[3]}. Finally, we describe the soft emission, which mainly arises from the extended regions, with a thermal diffuse emission, model {\tt mekal[8]}.

\begin{table*}
\begin{threeparttable}

\caption{Best-fit parameters obtained by fitting the spectra, using XSPEC, with Pexmon, MYTC, MYTD and Borus models. The values between brackets represent the mean value of each parameter and the corresponding 1$\sigma$ confidence interval obtained from the MCMC analysis. We also report the observed 2-10~keV fluxes and intrinsic luminosities for each model.}
\begin{tabular}{lllll}
\hline \hline
	Parameter			&	Pexmon	&	MYTC	&	MYTD	&	Borus	\\ \hline
$	N_{\rm H,\,LOS}\, (10^{24}\, \rm cm^{-2})		$	&	2.76 [3.43 (2.24, 4.45)]	&	--	&	4.67 [5.93 (3.86, 8.24)]	&	10 [5.81 (3.33, 8.51)]	\\
$	N_{\rm H,\,global} \, (10^{24}\, \rm cm^{-2})		$	&	--	&	10 [7.81 (6.17, 9.41)]	&	4.67$^t$	&	10$^t$	\\
$	\Gamma		$	&	2.4 [2.28 (2.18, 2.37)]	&	2.19 [2.20 (2.09, 2.31)]	&	2.33 [2.29 (2.22, 2.36)]	&	2.4 [2.11 (1.88, 2.33)]	\\
$	N_{\rm PL} \, (10^{-3})		$	&	2.46 [1.79 (1.33, 2.25)]	&	5.93 [9.81 (5.79, 13.99)]	&	6.61 [6.44 (5.15, 7.73)]	&	20.61 [5.49 (1.13, 9.91)]	\\
$	\theta_{\rm inc}		$	&	0$^f$	&	61.85 [68.77 (64.2, 73.3)]	&	--	&	63.38 [50.04 (34.7, 65)]	\\
$	\theta_{\rm torus}		$	&	--	&	--	&	--	&	60.65 [50.69 (38.2, 64.2)]	\\
$	\rm Abund		$	&	1.47 [1.40 (1.15, 1.62)]	&	--	&	--	&	1$^f$	\\
$	N_{\rm H,SC}\, (10^{22} \, \rm cm^{-2})		$	&	0.72 [0.88 (0.35, 1.36)]	&	0.25 [0.54 (0.19, 0.89)]	&	0.71 [0.76 (0.37, 1.13)]	&	0.72 [0.57 (0.27, 0.85)]	\\
$	f_{\rm SC} \, (10^{-3})		$	&	4.68 [9.22 (4.3, 13.3)]	&	1.39 [1.33 (0.64, 2.02)]	&	2.25 [2.58 (1.49, 3.67)]	&	0.93 [5.28 (1.44, 9.58)]	\\
$	kT_{\rm mekal}		$	&	0.68 [0.67 (0.62, 0.73)]	&	0.69 [0.67 (0.62, 0.74)]	&	0.67 [0.66 (0.61, 0.73)]	&	0.67$^f$	\\
$	N_{\rm mekal} \, ( 10^{-6})		$	&	4.56 [4.77 (3.86, 5.64)]	&	3.84 [4.45 (3.43, 5.43)]	&	4.70 [4.69 (3.78, 5.61)]	&	4.70$^f$	\\ \hline
$F_{2-10}\,(\rm 10^{-13}~erg~s^{-1}~cm^{-2})$ & $ 2.24^{+0.11}_{-0.07} $ & $2.34 \pm 0.11 $ & $2.32 ^{+0.10}_{-0.08}$ & $2.32 ^{+0.11}_{-0.07} $ \\ 
$L_{2-10}\,(\rm 10^{41}~erg~s^{-1})$ & $4.7 \pm 1.2$ & $15.2 \pm 7.6 $ & $14.4 \pm 2.9$ & $40.5 \pm 31.6$ \\ \hline
$	C/{\rm dof}		$	&	91.37/90	&	93.38/90	&	89.08/91	&	78.81/84	\\
\hline\hline
\end{tabular}
\begin{tablenotes}
	\item[$f$] fixed.
	\item[$t$] tied.
\end{tablenotes}
\end{threeparttable}
\label{table:param}
\end{table*}


The Pexmon model assumes an infinite slab responsible for the reprocessed emission. We were not able to constrain the inclination angle of the slab, so we fixed it to its best-fit value. We note that this angle represents the viewing angle of the reprocessing material itself and not of the whole system. We also fixed the reflection fraction to -1, so we account for the reflected emission only, assuming an isotropic primary emission. We left the Fe abundance (tied to the elemental abundance) free to vary.

This model fits the data well ($C/{\rm dof} = 91.37/90$). Throughout this work, we assess the goodness of the fits using simple Kolmogorov-Smirnov (KS) test in XSPEC, which estimates the largest difference ($\log D_{\rm KS}$) between the observed and model cumulative spectra. We also use the {\tt `goodness'} command in XSPEC by simulating 100 spectra based on the best-fit model and estimating the percentage of these simulations with the KS statistic less than that for the data (hereafter $p_{\rm KS}$). For this model, we found $\log D_{\rm KS} = -3.53$ and $p_{\rm KS} = 77\%$. The residuals are shown in Fig.~\ref{fig:spectra_model}c. The pexmon modelling implies that the emission in this source is absorbed by CT material with $N_{\rm H,\, LOS} = 2.76 \times 10^{24}\, \rm cm^{-2}$. Being heavily absorbed, the photon index of the primary emission could not be well constrained, so it was pegged to its maximum allowed value $\Gamma = 2.4$, with a lower limit of 1.95. The elemental abundance is found to be 1.45 times the solar abundance. We note that the scattered power-law emission is found to be $f_{\rm scat} = 0.47\%$ of the intrinsic primary emission, and absorbed by a material with $N_{\rm H, \, scat} =  7.2 \times 10^{21}\, \rm cm^{-2}$. The thermal mekal model adequately describes the soft X-rays with a temperature $kT = 0.68~ \rm keV$. The contour plots for the relevant parameters are shown in red in the lower panel of Fig.~\ref{fig:spectra_model}. Given the best-fit model, the intrinsic unabsorbed luminosity of the \textit{primary} emission is $L_{2-10} = (4.71 \pm 1.22)\times 10^{41}~~\rm erg\,s^{-1}$. The best-fit parameters for all models are listed in Table~\ref{table:param}.

\subsection{MYTorus}
\label{sec:MYT}

We next attempt to model the obscuration and the reprocessing emission by a CT torus using the MYTorus spectral-fitting suite for modeling X-ray spectra from a toroidal reprocessor \citep{MYT09}. We first consider the ``coupled'' configuration of MYTorus (hereafter MYTC). This configuration assumes that intrinsic emission is self-consistently absorbed and reprocessed by toroidal material with a circular cross section and half-opening angle of 60\degr\ and solar abundance. The viewing angle ($\theta_{\rm inc}$) and the global column density ($N_{\rm H,~global}$) of the torus are free parameters \citep[see][for more details about the various configurations of MYTorus]{Yaqoob12}. The model can be written as follows:

\begin{eqnarray}
\begin{array}{lll}
{\tt model_{MYTC} }  &=  {\tt phabs[1] *(  MYTZ[2] * zpowerlaw[3]  } &
\end{array}\nonumber\\
\begin{array}{lll}
  {\tt + zphabs[4]*constant[5]*zpowerlaw[6]  } & 
\end{array}\nonumber\\
\begin{array}{lll}
{\tt+ constant[7]*MYTS[8] + constant[9]*MYTL[10] }
\end{array}\nonumber \\
\begin{array}{lll}
{\tt + mekal[11]} ). &\nonumber
\end{array}
\label{eq:MYTC}
\end{eqnarray}
\noindent
The ${\tt phabs[1]}$, ${\tt zphabs[4]*constant[5]*zpowerlaw[6]}$ and {\tt mekal[11]} components are equivalent to the ones in the Pexmon fit. {\tt MYTZ[2]} represents the attenuation of the intrinsic emission. {\tt MYTS[8]} and {\tt MYTL[10]} represent the scattered continuum and the fluorescent emission lines emitted by the torus. The {\tt constant[7,9]} factors correspond the relative weights of the three MYTorus components and are fixed to unity \citep[as suggested by][]{Yaqoob12}. We tried to link {\tt constant[7]} and  {\tt constant[9]} leaving the former free to vary. The quality of the fit was the same and we could not get any constraints on that parameter. We remind the reader that $N_{\rm H, LOS}$ can be estimated using $N_{\rm H,\, global}$ and $\theta_{\rm inc}$ by using eq. (1) in \cite{MYT09}. MYTorus does not have a high-energy cutoff. Imposing a cutoff energy to the primary emission would break the self-consistency of the models. Instead, MYTorus assumes various termination energies ($E_{\rm T}$). We used in our analysis the tables with $E_{\rm T} =500$~keV. Using different values did not affect the fits. 

MYTC provides a statistically acceptable fit ($C/\rm dof = 93.38/90$, $\log D_{\rm KS} = -3.4$, $ p_{\rm KS} = 80\%$). The residuals are shown in Fig.~\ref{fig:spectra_model}d. This model also implies a CT absorber with $N_{\rm H, \, global} $ pegged to its maximum allowed limit of $10^{25} \, \rm cm^{-2}$ (the lower limit is $4.34\times 10^{24}\, \rm cm^{-2}$) and $\theta_{\rm inc}\sim 62\degr$. This suggests $N_{\rm H,LOS} \sim 3.4 \times 10^{24}\, \rm cm^{-2}$, consistent with the value obtained from the Pexmon fit. We found that $\theta_{\rm inc}$ is close to the half-opening angle of the torus. This implies that, in the context of a toroidal geometry, a considerable contribution to the reprocessed emission comes from the far side of the torus.

Next, we considered the decoupled configuration of MYTorus (hereafter MYTD) which is intended to mimic the Pexmon configuration. In this configuration, the viewing angle of {\tt MYTZ} is fixed to 90\degr, so its $N_{\rm H}$ corresponds to the LOS value. {\tt MYTS} and {\tt MYTL} are decomposed into two components, one from the near side of the torus ($\theta_{\rm inc}= 90\degr$) and the one from the far side of the torus ($\theta_{\rm inc} = 0 \degr$). The column densities of these component could be either tied to the one of {\tt MYTZ}, corresponding to a uniform distribution of the material, or free to vary (corresponding to a patchy structure). This model has only two more free parameters with respect to MYTC, which are the weights of {\tt MYTS} and {\tt MYTL}. MYTD can be written as follows:

\begin{eqnarray}
\begin{array}{lll}
{\tt model_{MYTD} }  &=  {\tt phabs[1] *(  MYTZ_{90}[2] * zpowerlaw[3]  } &
\end{array}\nonumber\\
\begin{array}{lll}
  {\tt + zphabs[4]*constant[5]*zpowerlaw[6]  } & 
\end{array}\nonumber\\
\begin{array}{lll}
{\tt + constant[7]*(MYTS_0[8] + MYTL_0[9]) }
\end{array}\nonumber \\
\begin{array}{lll}
{\tt+ constant[10]*(MYTS_{90}[11] + MYTL_{90}[12]) }
\end{array}\nonumber \\
\begin{array}{lll}
{\tt + mekal[13]} ). &\nonumber
\end{array}
\label{eq:MYTD}
\end{eqnarray}
\noindent
For this configuration, we kept the column densities for all the {\tt MYT} components tied to the LOS value. Letting it be free resulted in a similar result. The relative weights for the {\tt MYTS} and {\tt MYTL} components with the same $\theta_{\rm inc}$ are tied together. First, we left {\tt constant[10]} free to vary. However, as expected from a CT absorption in the LOS, the reprocessed emission would be unlikely to escape the near side of the torus. Indeed, we find {\tt constant[11]} to be negligible ($< 10^{-5}$). Hence, we fixed it to zero in the rest of the analysis. We also fixed {\tt constant[7]} to unity which reduces the number of free parameters. We could not constrain it by leaving it free to vary. The model provides a statistically acceptable fit ($C/\rm dof = 89.08/91$, $\log D_{\rm KS} = -3.5$, $p_{\rm KS} = 79\%$). The best-fit parameters are listed in Table~\ref{table:param} and the residuals for this model are shown in Fig.~\ref{fig:spectra_model}e. The various components for this model are shown in panel (a) of the same figure. This figure shows clearly that the observed spectrum is dominated above $\sim 2$ keV by the reprocessed emission from CT material ($N_{\rm H,\, LOS} = 4.67 \times 10^{24}~\rm cm^{-2}$). The contour plots are shown in blue in the lower panel of Fig. \ref{fig:spectra_model}. All the parameters are consistent with the ones given by the Pexmon fit, except the normalization of the primary emission which is $\sim 2.7$ times larger than the value inferred by Pexmon. This is mainly due to the fact that the primary emission cannot be seen due to obscuration (see the red dotted line in Fig.~\ref{fig:spectra_model}a). Thus its intrinsic luminosity is estimated indirectly based on the reprocessed emission. This may lead to the discrepancy in the normalization of the primary due to different physical assumptions. Given the best-fit model, the observed 2-10~keV and 10-30~keV fluxes of this source are $(2.32 \pm 0.1)\times 10^{-13}\rm ~erg\,s^{-1}$ and $(1.16\pm 0.08)\times 10^{-12} \rm ~erg\,s^{-1}~cm^{-2}$, respectively. The intrinsic unabsorbed luminosity of the \textit{primary} emission is $L_{2-10} = (1.44 \pm 0.29)\times 10^{42}~\rm erg\,s^{-1}~cm^{-2}$. Thus, the intrinsic to observed flux ratio at $2-10$~keV is $103 \pm 23$.

\subsection{Borus}
\label{sec:borus}

Finally, we fit the spectra of the source by modeling the reprocessed emission using the Borus model \citep{Borus18}. Borus assumes a uniform density sphere with cutouts which are determined by the half-opening angle $\theta_{\rm torus}$ (a free parameter), giving a similar geometry to the one of MYTorus. The model can be described as follows:
\begin{eqnarray}
\begin{array}{lll}
{\tt model_{Borus} }  &=  {\tt phabs[1] *(  zphabs[2]*cabs[3] * cutoffpl[4]  } &
\end{array}\nonumber\\
\begin{array}{lll}
  {\tt + zphabs[5]*constant[6]*cutoffpl[7]  + Borus[8] } & 
\end{array}\nonumber\\
\begin{array}{lll}
{\tt +  mekal[9]} ). &\nonumber
\end{array}
\label{eq:borus}
\end{eqnarray}
\noindent
The {\tt cabs[3]} component accounts for Compton scattering in the absorber, while the {\tt Borus[8]} accounts for the reprocessed emission. We note that the {\tt Borus} model is defined above 1~keV. For that reason, we fixed the parameters of the {\tt mekal[9]} component to the best-fit values obtained by the MYTD model and fitted the spectra above 1~keV. We tied $N_{\rm H,LOS}$ to the global value. We also tied the normalization of {\tt Borus[8]} to the one of {\tt cutoffpl[4]} whose high-energy cutoff is fixed to 500~keV. The abundance could not be constrained, so we fixed it to the solar value. The model gives a statistically acceptable fit ($C/\rm dof = 78.81/84$, $\log D_{\rm KS} = -3.5$, $p_{\rm KS} = 66\%$), also implying a CT source ($N_{\rm H,~ LOS} = 10^{25}\rm cm^{-2}$). We note that all the parameters are consistent with the values obtained by MYTD. Notably, $\theta_{\rm torus} \sim 60.4\degr$ is consistent with the one assumed by construction in the MYTorus model, which is 60\degr. The intrinsic unabsorbed luminosity of the \textit{primary} emission is $L_{2-10} = (4.05 \pm 3.16)\times 10^{42}~\rm erg\,s^{-1}$, consistent with the value obtained from the MYTD model.

\section{Discussion and conclusions}
\label{sec:discussion}

\begin{figure*}
\centering
\includegraphics[width = 0.49\linewidth]{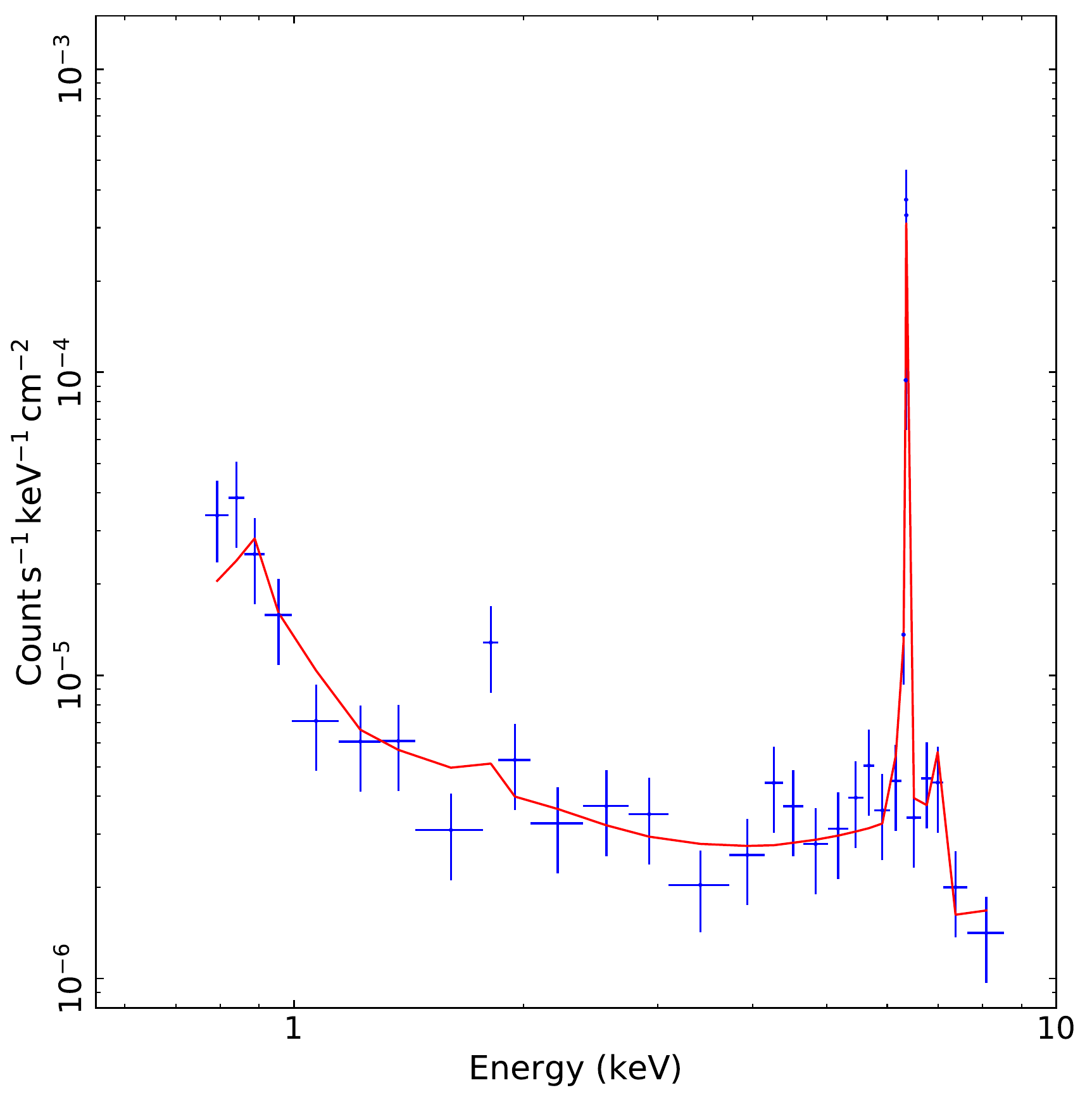}
\includegraphics[width = 0.49\linewidth]{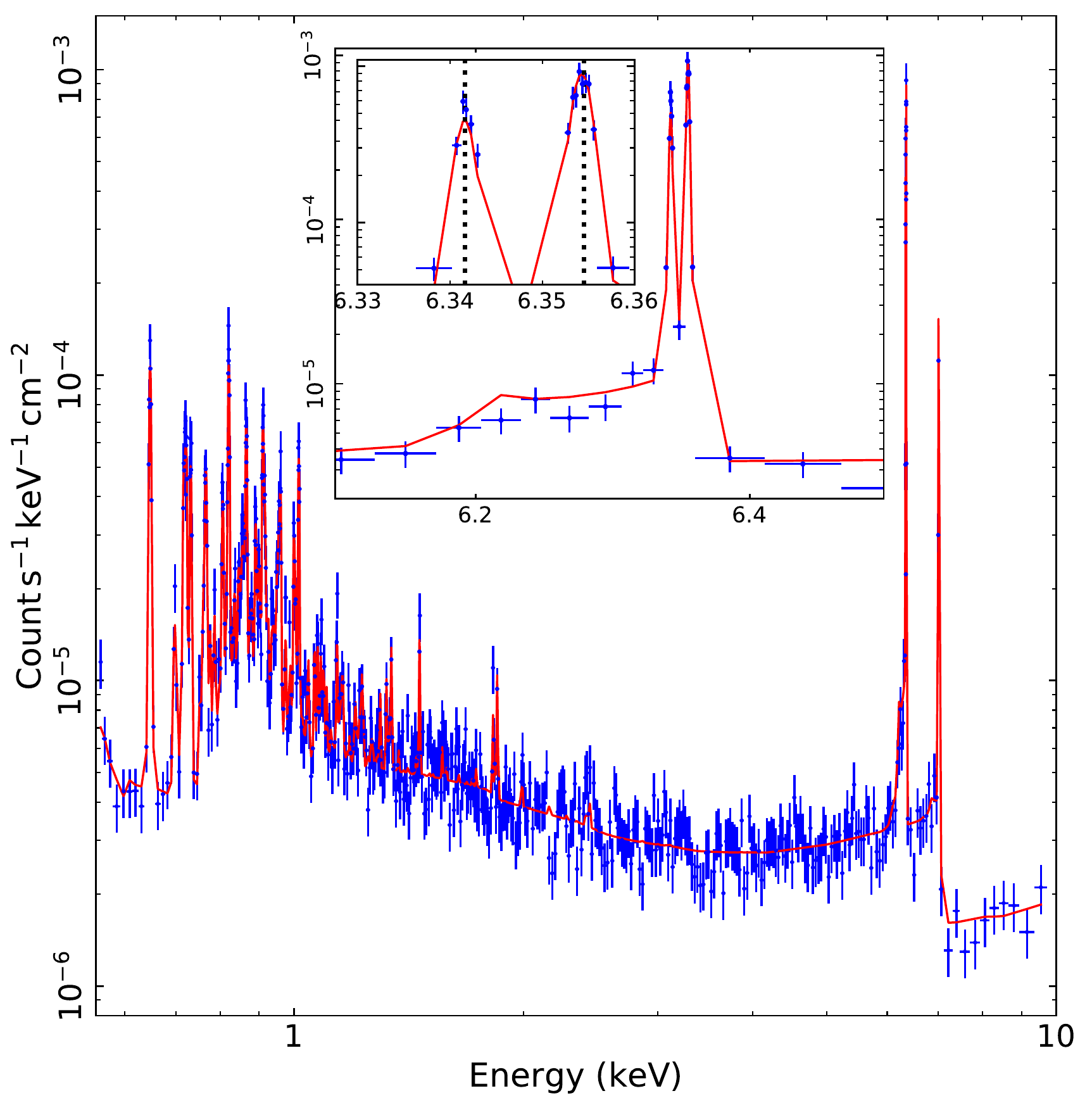} 
\caption{Simulated \textit{XRISM} (left panel) and \textit{Athena}/X-IFU (right panel) spectra of NGC 5347 assuming the best-fit MYTD model (red line) and an exposure time of 100 ks(in the observed frame). The inset in the right panel shows a zoom in on the 6--6.5~keV range, where the Fe K$\alpha$1,2 lines can be clearly resolved and separated together with the Compton shoulder. The vertical dotted lines show the fiducial energies of the Fe K$\alpha$1,2 lines assumed to be at 6.404~keV and 6.391~keV (rest frame), respectively. The spectra were grouped to require at least 10 and 25 counts per bin, for \textit{XRISM} and \textit{Athena}, respectively.}
\label{fig:Xrism-athena}
\end{figure*}

We have confirmed in this work the CT classification of NGC 5347. The $2-30$~keV multi-epoch spectra of this source are clearly dominated by reprocessed emission from CT material ($N_{\rm H} > 2.23 \times 10^{24}~\rm cm^{-2}$) obscuring the central engine. We note that this estimate is higher than the previously reported ones for this source \citep{Risaliti99,Lamassa11}. It is then possible that some sources which were classified as Compton thin could be in reality CT when higher-quality data covering a broader energy range and more physical models are used. \cite{Marchesi18} estimated $N_{\rm H}$ for 30 local AGN ($\langle z \rangle \sim 0.03$) from the \textit{Swift}/BAT 100-month survey that were observed by \textit{NuSTAR}. Our estimate puts the source in the upper quartile of the $N_{\rm H}$-distribution presented in Fig. 3 of \cite{Marchesi18}.

It is worth mentioning that all the models employed in this work are statistically comparable, giving statistically good fits and consistent physical parameters. However, the use of physically-motivated models such as MYTorus and Borus, accounting properly for the reprocessed emission in the torus, is preferred with respect to simple reflection models. We note that due to the faintness of the source, we were not able to get strong constraints either on the geometry of the obscuring material or on the properties of the intrinsic X-ray source. Our results are in alignment with the studies of megamaser AGN \citep[e.g.,][]{Greenhill08, Masini16} which revealed that a large fraction of megamaser AGN harbor a CTAGN.

Using the unabsorbed $L_{2-10}$ nuclear emission from the best-fit MYTD model, we solved the third-degree equation provided by \cite{Marconi04} (in their equation 21) to estimate the bolometric luminosity ($L_{\rm bol}$). We obtain $L_{\rm bol} = (1.65 \pm 0.33) \times 10^{43}~{\rm erg\,s^{-1}} \simeq 0.014 \pm 0.005~L_{\rm Edd}$. Moreover, by applying the $L_{2-10}-L_{\rm [O~III]}$ relationship for Seyferts found by \cite{Berney15}, we obtain $\log \left( L_{\rm [O~III]}/\rm erg~s^{-1} \right) \sim 40.14$ which is consistent with the values we obtained from the SDSS spectrum (see Fig.~\ref{fig:sdss}; $\log \left(L_{\rm [O~III]}^{\rm int.}/\rm erg~s^{-1} \right) = 40.11 \pm 0.03$) and the one estimated by \cite{Schmitt03} from \textit{HST} observations of this source. NGC 5347 is therefore consistent with the other sources analyzed by \cite{Ueda15} in both the $\log N_{\rm H} - \log \left[ L_{\rm [O~III]}/L_{2-10} \right] $ and the $\log \left[ L_{2-10}/L_{\rm Edd}\right] - \log \left[ L_{\rm [O~III]}/L_{2-10}\right]$ planes. However, the observed [\ion{O}{4}] luminosity \citep[$L_{\rm [O~IV]} = 1.15 \times 10^{40}~\rm erg~s^{-1}$;][]{wu11} is smaller than the inferred value ($L_{\rm [O~IV]} = 7.24 \times 10^{40}~\rm erg~s^{-1}$) obtained from $L_{\rm [O~IV]}-L_{2-10}$ correlation by \citep{Melendez08}. The two values are broadly consistent given the large uncertainties in this correlation. We note that the measured [\ion{O}{3}] and [\ion{O}{4}] fluxes could be underestimated if the NLR extends beyond the slit size of the instruments due to the closeness of the source. Finally, following \cite{Risaliti11} and \cite{Bis16}, if we assume that the [\ion{O}{3}] luminosity is an indicator of the intrinsic luminosity and is emitted isotropically, and that the underlying continuum is due to an optically thick accretion disk, then the observed [\ion{O}{3}] EW (EW$_{\rm obs}$) can give us an estimate of the orientation of the accretion disk. For an accretion disk that is observed with an inclination $\theta$, we get ${\rm EW_{\rm obs} = EW^\ast}/\cos \theta$, where ${\rm EW^\ast}$ is the EW as measured in a face-on configuration. Ideally the latter value is the same for all AGN. By considering an average value of $\rm EW^\ast$ to be $\sim 11~\AA$ \citep[see][who estimated $\langle \rm EW^\ast \rangle$ for a sample of SDSS quasars]{Risaliti11, Bis16}, and $\rm EW_{obs} = 16.91~\AA$ (see Table~\ref{table:optlines}), we obtain $\theta \sim 50\degr$.

Given the quality of the current data, we modeled the soft X-ray spectra ($E \leq 2~\rm keV$) using a fiducial model which assumes thermal diffuse emission. This model ensured a fair representation of the observed spectra at these energies. However, some of the emission could be due to photoionized plasma in the NLR. Understanding the nature of the soft X-ray emission in this source requires higher-quality data, which makes it an interesting candidate for future planned X-ray observatories carrying micro-calorimeters, such as \textit{XRISM} and \textit{Athena} \citep{Athena13}. 

Future X-ray missions will allow us to obtain higher quality spectra by resolving all the line features in the spectra, enabling a better understanding of such sources. Fig.~\ref{fig:Xrism-athena} shows 100-ks simulated spectra using the response files of \textit{XRISM} (left panel) and \textit{Athena} /X-IFU\footnote{\url{http://x-ifu.irap.omp.eu/resources-for-users-and-x-ifu-consortium-members/}} \citep[right panel;][]{xifu13}, assuming the best-fit MYTD model. The faintness of the source would not allow the emission lines to be well resolved using XRISM. However, thanks to the large effective area of \textit{Athena}  all emission lines would be easily resolved. More interestingly, the inset in the right panel of this figure shows clearly that both Fe K$\alpha$1,2 lines (at rest-frame energies of 6.404~keV and 6.391~keV, respectively) would be separated and resolved, in addition to the corresponding Compton shoulder. This will enable a better identification and characterization of faint CT sources, with high-quality spectra. It will also allow us to study and identify the various spectral components in these sources, the geometry and covering fraction of the obscuring material. Any motion of the absorbing/reprocessing material whether it is orbital and/or outflowing will be then imprinted in the lines profile in terms of broadening and/or energy shift, and will be easily identified. Moreover, the rich emission line spectrum in the soft X-rays will allow us to study the various components contributing to this energy range such as the thermal emission from the host galaxy, or the extended emission from the NLR.

\begin{acknowledgements}
KO is an International Research Fellow of the Japan Society for the Promotion of Science (JSPS) (ID: P17321). We would like to thank Karl Forster for scheduling all of  the \textit{NuSTAR} observations in our program. We would like to thank Xavier Barcons, Fiona Harrison, Tim Kallman, Kirpal Nandra and John Raymond for useful conversations. This work made use of data from the NuSTAR mission, a project led by the California Institute of Technology, managed by the Jet Propulsion Laboratory, and funded by the National Aeronautics and Space Administration.The results presented in this paper are also based on data obtained with the {\it Suzaku} observatory; and the {\it Chandra} X-ray Observatory. The figures were generated using matplotlib \citep{Hunter07}, a {\tt PYTHON} library for publication of quality graphics. The MCMC results were presented using the GetDist {\tt PYTHON} package.
\end{acknowledgements}

\software{pPXF \citep{Cappellari04}, CIAO \citep[v9.4][]{Ciao}, HEASoft \citep{HEASoft}, NUSTARDAS (v1.8.0, \url{https://heasarc.gsfc.nasa.gov/docs/nustar/analysis/}, XSPEC \citep{Arnaud96}, Matplotlib \citep{Hunter07}, GetDist (\url{https://getdist.readthedocs.io/en/latest/})}

\facilities{\textit{Chanrda X-ray Observatory}, SDSS, {\it NuSTAR, Suzaku.}}

\begin{thebibliography}{}
\expandafter\ifx\csname natexlab\endcsname\relax\def\natexlab#1{#1}\fi
\providecommand{\url}[1]{\href{#1}{#1}}
\providecommand{\dodoi}[1]{doi:~\href{http://doi.org/#1}{\nolinkurl{#1}}}
\providecommand{\doeprint}[1]{\href{http://ascl.net/#1}{\nolinkurl{http://ascl.net/#1}}}
\providecommand{\doarXiv}[1]{\href{https://arxiv.org/abs/#1}{\nolinkurl{https://arxiv.org/abs/#1}}}

\bibitem[{{Abazajian} {et~al.}(2009){Abazajian}, {Adelman-McCarthy},
  {Ag{\"u}eros}, {Allam}, {Allende Prieto}, {An}, {Anderson}, {Anderson},
  {Annis}, {Bahcall}, \& et~al.}]{SDSSDR7}
{Abazajian}, K.~N., {Adelman-McCarthy}, J.~K., {Ag{\"u}eros}, M.~A., {et~al.}
  2009, \apjs, 182, 543, \dodoi{10.1088/0067-0049/182/2/543}

\bibitem[{{Antonucci}(1993)}]{Antonucci93}
{Antonucci}, R. 1993, \araa, 31, 473,
  \dodoi{10.1146/annurev.aa.31.090193.002353}

\bibitem[{{Arnaud}(1996)}]{Arnaud96}
{Arnaud}, K.~A. 1996, in Astronomical Society of the Pacific Conference Series,
  Vol. 101, Astronomical Data Analysis Software and Systems V, ed. G.~H.
  {Jacoby} \& J.~{Barnes}, 17

\bibitem[{{Balokovi{\'c}} {et~al.}(2014){Balokovi{\'c}}, {Comastri},
  {Harrison}, {Alexander}, {Ballantyne}, {Bauer}, {Boggs}, {Brandt},
  {Brightman}, {Christensen}, {Craig}, {Del Moro}, {Gandhi}, {Hailey}, {Koss},
  {Lansbury}, {Luo}, {Madejski}, {Marinucci}, {Matt}, {Markwardt}, {Puccetti},
  {Reynolds}, {Risaliti}, {Rivers}, {Stern}, {Walton}, \&
  {Zhang}}]{Balokovic14}
{Balokovi{\'c}}, M., {Comastri}, A., {Harrison}, F.~A., {et~al.} 2014, \apj,
  794, 111, \dodoi{10.1088/0004-637X/794/2/111}

\bibitem[{{Balokovi{\'c}} {et~al.}(2018){Balokovi{\'c}}, {Brightman},
  {Harrison}, {Comastri}, {Ricci}, {Buchner}, {Gandhi}, {Farrah}, \&
  {Stern}}]{Borus18}
{Balokovi{\'c}}, M., {Brightman}, M., {Harrison}, F.~A., {et~al.} 2018, \apj,
  854, 42, \dodoi{10.3847/1538-4357/aaa7eb}

\bibitem[{{Barret} {et~al.}(2013){Barret}, {den Herder}, {Piro}, {Ravera}, {Den
  Hartog}, {Macculi}, {Barcons}, {Page}, {Paltani}, {Rauw}, {Wilms},
  {Ceballos}, {Duband}, {Gottardi}, {Lotti}, {de Plaa}, {Pointecouteau},
  {Schmid}, {Akamatsu}, {Bagliani}, {Bandler}, {Barbera}, {Bastia}, {Biasotti},
  {Branco}, {Camon}, {Cara}, {Cobo}, {Colasanti}, {Costa-Kramer}, {Corcione},
  {Doriese}, {Duval}, {Fabrega}, {Gatti}, {de Gerone}, {Guttridge}, {Kelley},
  {Kilbourne}, {van der Kuur}, {Mineo}, {Mitsuda}, {Natalucci}, {Ohashi},
  {Peille}, {Perinati}, {Pigot}, {Pizzigoni}, {Pobes}, {Porter}, {Renotte},
  {Sauvageot}, {Sciortino}, {Torrioli}, {Valenziano}, {Willingale}, {de Vries},
  \& {van Weers}}]{xifu13}
{Barret}, D., {den Herder}, J.~W., {Piro}, L., {et~al.} 2013, ArXiv e-prints.
\newblock \doarXiv{1308.6784}

\bibitem[{{Berney} {et~al.}(2015){Berney}, {Koss}, {Trakhtenbrot}, {Ricci},
  {Lamperti}, {Schawinski}, {Balokovi{\'c}}, {Crenshaw}, {Fischer}, {Gehrels},
  {Harrison}, {Hashimoto}, {Ichikawa}, {Mushotzky}, {Oh}, {Stern}, {Treister},
  {Ueda}, {Veilleux}, \& {Winter}}]{Berney15}
{Berney}, S., {Koss}, M., {Trakhtenbrot}, B., {et~al.} 2015, \mnras, 454, 3622,
  \dodoi{10.1093/mnras/stv2181}

\bibitem[{{Bisogni} {et~al.}(2016){Bisogni}, {Marconi}, \& {Risaliti}}]{Bis16}
{Bisogni}, S., {Marconi}, A., \& {Risaliti}, G. 2016, \mnras,
  \dodoi{10.1093/mnras/stw2324}

\bibitem[{{Brightman} {et~al.}(2018){Brightman}, {Balokovi{\'c}}, {Koss},
  {Alexander}, {Annuar}, {Earnshaw}, {Gandhi}, {Harrison}, {Hornschemeier},
  {Lehmer}, {Powell}, {Ptak}, {Rangelov}, {Roberts}, {Stern}, {Walton}, \&
  {Zezas}}]{Brightman18}
{Brightman}, M., {Balokovi{\'c}}, M., {Koss}, M., {et~al.} 2018, \apj, 867,
  110, \dodoi{10.3847/1538-4357/aae1ae}

\bibitem[{{Bruzual} \& {Charlot}(2003)}]{Bruzual03}
{Bruzual}, G., \& {Charlot}, S. 2003, \mnras, 344, 1000,
  \dodoi{10.1046/j.1365-8711.2003.06897.x}

\bibitem[{{Cappellari} \& {Emsellem}(2004)}]{Cappellari04}
{Cappellari}, M., \& {Emsellem}, E. 2004, \pasp, 116, 138,
  \dodoi{10.1086/381875}

\bibitem[{{Cardelli} {et~al.}(1989){Cardelli}, {Clayton}, \&
  {Mathis}}]{Cardelli89}
{Cardelli}, J.~A., {Clayton}, G.~C., \& {Mathis}, J.~S. 1989, \apj, 345, 245,
  \dodoi{10.1086/167900}

\bibitem[{{Cash}(1979)}]{Cash79}
{Cash}, W. 1979, \apj, 228, 939, \dodoi{10.1086/156922}

\bibitem[{{Chen} {et~al.}(2014){Chen}, {Trager}, {Peletier}, {Lan{\c c}on},
  {Vazdekis}, {Prugniel}, {Silva}, \& {Gonneau}}]{Chen14}
{Chen}, Y.-P., {Trager}, S.~C., {Peletier}, R.~F., {et~al.} 2014, \aap, 565,
  A117, \dodoi{10.1051/0004-6361/201322505}

\bibitem[{{de Grijp} {et~al.}(1992){de Grijp}, {Keel}, {Miley}, {Goudfrooij},
  \& {Lub}}]{degrijp92}
{de Grijp}, M.~H.~K., {Keel}, W.~C., {Miley}, G.~K., {Goudfrooij}, P., \&
  {Lub}, J. 1992, \aaps, 96, 389

\bibitem[{{Fruscione} {et~al.}(2006){Fruscione}, {McDowell}, {Allen},
  {Brickhouse}, {Burke}, {Davis}, {Durham}, {Elvis}, {Galle}, {Harris},
  {Huenemoerder}, {Houck}, {Ishibashi}, {Karovska}, {Nicastro}, {Noble},
  {Nowak}, {Primini}, {Siemiginowska}, {Smith}, \& {Wise}}]{Ciao}
{Fruscione}, A., {McDowell}, J.~C., {Allen}, G.~E., {et~al.} 2006, in
  \procspie, Vol. 6270, Society of Photo-Optical Instrumentation Engineers
  (SPIE) Conference Series, 62701V

\bibitem[{{Goodman} \& {Weare}(2010)}]{Goodman10}
{Goodman}, J., \& {Weare}, J. 2010, Comm. App. Math. Comp. Sci., 5, 65,
  \dodoi{10.2140/camcos.2010.5.65}

\bibitem[{{Greenhill} {et~al.}(2008){Greenhill}, {Tilak}, \&
  {Madejski}}]{Greenhill08}
{Greenhill}, L.~J., {Tilak}, A., \& {Madejski}, G. 2008, \apjl, 686, L13,
  \dodoi{10.1086/592782}

\bibitem[{{Harrison} {et~al.}(2013){Harrison}, {Craig}, {Christensen},
  {Hailey}, {Zhang}, {Boggs}, {Stern}, {Cook}, {Forster}, {Giommi},
  {Grefenstette}, {Kim}, {Kitaguchi}, {Koglin}, {Madsen}, {Mao}, {Miyasaka},
  {Mori}, {Perri}, {Pivovaroff}, {Puccetti}, {Rana}, {Westergaard}, {Willis},
  {Zoglauer}, {An}, {Bachetti}, {Barri{\`e}re}, {Bellm}, {Bhalerao},
  {Brejnholt}, {Fuerst}, {Liebe}, {Markwardt}, {Nynka}, {Vogel}, {Walton},
  {Wik}, {Alexander}, {Cominsky}, {Hornschemeier}, {Hornstrup}, {Kaspi},
  {Madejski}, {Matt}, {Molendi}, {Smith}, {Tomsick}, {Ajello}, {Ballantyne},
  {Balokovi{\'c}}, {Barret}, {Bauer}, {Blandford}, {Brandt}, {Brenneman},
  {Chiang}, {Chakrabarty}, {Chenevez}, {Comastri}, {Dufour}, {Elvis}, {Fabian},
  {Farrah}, {Fryer}, {Gotthelf}, {Grindlay}, {Helfand}, {Krivonos}, {Meier},
  {Miller}, {Natalucci}, {Ogle}, {Ofek}, {Ptak}, {Reynolds}, {Rigby},
  {Tagliaferri}, {Thorsett}, {Treister}, \& {Urry}}]{Harrison13}
{Harrison}, F.~A., {Craig}, W.~W., {Christensen}, F.~E., {et~al.} 2013, \apj,
  770, 103, \dodoi{10.1088/0004-637X/770/2/103}

\bibitem[{{Hern{\'a}n-Caballero} {et~al.}(2015){Hern{\'a}n-Caballero},
  {Alonso-Herrero}, {Hatziminaoglou}, {Spoon}, {Ramos Almeida}, {D{\'{\i}}az
  Santos}, {H{\"o}nig}, {Gonz{\'a}lez-Mart{\'{\i}}n}, \& {Esquej}}]{Hernan15}
{Hern{\'a}n-Caballero}, A., {Alonso-Herrero}, A., {Hatziminaoglou}, E.,
  {et~al.} 2015, \apj, 803, 109, \dodoi{10.1088/0004-637X/803/2/109}

\bibitem[{{H{\"o}nig} \& {Beckert}(2007)}]{Honig07}
{H{\"o}nig}, S.~F., \& {Beckert}, T. 2007, \mnras, 380, 1172,
  \dodoi{10.1111/j.1365-2966.2007.12157.x}

\bibitem[{{Huchra} {et~al.}(1983){Huchra}, {Davis}, {Latham}, \&
  {Tonry}}]{Huchra83}
{Huchra}, J., {Davis}, M., {Latham}, D., \& {Tonry}, J. 1983, \apjs, 52, 89,
  \dodoi{10.1086/190860}

\bibitem[{Hunter(2007)}]{Hunter07}
Hunter, J.~D. 2007, Computing In Science \& Engineering, 9, 90,
  \dodoi{10.1109/MCSE.2007.55}

\bibitem[{{Izumi} {et~al.}(2016){Izumi}, {Kawakatu}, \& {Kohno}}]{Izumi16}
{Izumi}, T., {Kawakatu}, N., \& {Kohno}, K. 2016, \apj, 827, 81,
  \dodoi{10.3847/0004-637X/827/1/81}

\bibitem[{{Kalberla} {et~al.}(2005){Kalberla}, {Burton}, {Hartmann}, {Arnal},
  {Bajaja}, {Morras}, \& {P{\"o}ppel}}]{Kalberla05}
{Kalberla}, P.~M.~W., {Burton}, W.~B., {Hartmann}, D., {et~al.} 2005, \aap,
  440, 775, \dodoi{10.1051/0004-6361:20041864}

\bibitem[{{Kewley} {et~al.}(2006){Kewley}, {Groves}, {Kauffmann}, \&
  {Heckman}}]{Kewley06}
{Kewley}, L.~J., {Groves}, B., {Kauffmann}, G., \& {Heckman}, T. 2006, \mnras,
  372, 961, \dodoi{10.1111/j.1365-2966.2006.10859.x}

\bibitem[{{Kormendy} \& {Ho}(2013)}]{Kormendy13}
{Kormendy}, J., \& {Ho}, L.~C. 2013, \araa, 51, 511,
  \dodoi{10.1146/annurev-astro-082708-101811}

\bibitem[{{Koss} {et~al.}(2017){Koss}, {Trakhtenbrot}, {Ricci}, {Lamperti},
  {Oh}, {Berney}, {Schawinski}, {Balokovi{\'c}}, {Baronchelli}, {Crenshaw},
  {Fischer}, {Gehrels}, {Harrison}, {Hashimoto}, {Hogg}, {Ichikawa}, {Masetti},
  {Mushotzky}, {Sartori}, {Stern}, {Treister}, {Ueda}, {Veilleux}, \&
  {Winter}}]{Koss17}
{Koss}, M., {Trakhtenbrot}, B., {Ricci}, C., {et~al.} 2017, \apj, 850, 74,
  \dodoi{10.3847/1538-4357/aa8ec9}

\bibitem[{{Koss} {et~al.}(2016){Koss}, {Glidden}, {Balokovi{\'c}}, {Stern},
  {Lamperti}, {Assef}, {Bauer}, {Ballantyne}, {Boggs}, {Craig}, {Farrah},
  {F{\"u}rst}, {Gandhi}, {Gehrels}, {Hailey}, {Harrison}, {Markwardt},
  {Masini}, {Ricci}, {Treister}, {Walton}, \& {Zhang}}]{Koss16}
{Koss}, M.~J., {Glidden}, A., {Balokovi{\'c}}, M., {et~al.} 2016, \apjl, 824,
  L4, \dodoi{10.3847/2041-8205/824/1/L4}

\bibitem[{{Koyama} {et~al.}(2007){Koyama}, {Tsunemi}, {Dotani}, {Bautz},
  {Hayashida}, {Tsuru}, {Matsumoto}, {Ogawara}, {Ricker}, {Doty}, {Kissel},
  {Foster}, {Nakajima}, {Yamaguchi}, {Mori}, {Sakano}, {Hamaguchi},
  {Nishiuchi}, {Miyata}, {Torii}, {Namiki}, {Katsuda}, {Matsuura}, {Miyauchi},
  {Anabuki}, {Tawa}, {Ozaki}, {Murakami}, {Maeda}, {Ichikawa}, {Prigozhin},
  {Boughan}, {Lamarr}, {Miller}, {Burke}, {Gregory}, {Pillsbury}, {Bamba},
  {Hiraga}, {Senda}, {Katayama}, {Kitamoto}, {Tsujimoto}, {Kohmura}, {Tsuboi},
  \& {Awaki}}]{XIS}
{Koyama}, K., {Tsunemi}, H., {Dotani}, T., {et~al.} 2007, \pasj, 59, 23,
  \dodoi{10.1093/pasj/59.sp1.S23}

\bibitem[{{LaMassa} {et~al.}(2011){LaMassa}, {Heckman}, {Ptak}, {Martins},
  {Wild}, {Sonnentrucker}, \& {Hornschemeier}}]{Lamassa11}
{LaMassa}, S.~M., {Heckman}, T.~M., {Ptak}, A., {et~al.} 2011, \apj, 729, 52,
  \dodoi{10.1088/0004-637X/729/1/52}

\bibitem[{{Levenson} {et~al.}(2006){Levenson}, {Heckman}, {Krolik}, {Weaver},
  \& {{\.Z}ycki}}]{Levenson06}
{Levenson}, N.~A., {Heckman}, T.~M., {Krolik}, J.~H., {Weaver}, K.~A., \&
  {{\.Z}ycki}, P.~T. 2006, \apj, 648, 111, \dodoi{10.1086/505735}

\bibitem[{{Marchesi} {et~al.}(2018){Marchesi}, {Ajello}, {Marcotulli},
  {Comastri}, {Lanzuisi}, \& {Vignali}}]{Marchesi18}
{Marchesi}, S., {Ajello}, M., {Marcotulli}, L., {et~al.} 2018, \apj, 854, 49,
  \dodoi{10.3847/1538-4357/aaa410}

\bibitem[{{Marconi} {et~al.}(2004){Marconi}, {Risaliti}, {Gilli}, {Hunt},
  {Maiolino}, \& {Salvati}}]{Marconi04}
{Marconi}, A., {Risaliti}, G., {Gilli}, R., {et~al.} 2004, \mnras, 351, 169,
  \dodoi{10.1111/j.1365-2966.2004.07765.x}

\bibitem[{{Marinucci} {et~al.}(2016){Marinucci}, {Bianchi}, {Matt},
  {Alexander}, {Balokovi{\'c}}, {Bauer}, {Brandt}, {Gandhi}, {Guainazzi},
  {Harrison}, {Iwasawa}, {Koss}, {Madsen}, {Nicastro}, {Puccetti}, {Ricci},
  {Stern}, \& {Walton}}]{Marinucci16}
{Marinucci}, A., {Bianchi}, S., {Matt}, G., {et~al.} 2016, \mnras, 456, L94,
  \dodoi{10.1093/mnrasl/slv178}

\bibitem[{{Masini} {et~al.}(2016){Masini}, {Comastri}, {Balokovi{\'c}}, {Zaw},
  {Puccetti}, {Ballantyne}, {Bauer}, {Boggs}, {Brandt}, {Brightman},
  {Christensen}, {Craig}, {Gandhi}, {Hailey}, {Harrison}, {Koss}, {Madejski},
  {Ricci}, {Rivers}, {Stern}, \& {Zhang}}]{Masini16}
{Masini}, A., {Comastri}, A., {Balokovi{\'c}}, M., {et~al.} 2016, \aap, 589,
  A59, \dodoi{10.1051/0004-6361/201527689}

\bibitem[{{Mel{\'e}ndez} {et~al.}(2008){Mel{\'e}ndez}, {Kraemer}, {Armentrout},
  {Deo}, {Crenshaw}, {Schmitt}, {Mushotzky}, {Tueller}, {Markwardt}, \&
  {Winter}}]{Melendez08}
{Mel{\'e}ndez}, M., {Kraemer}, S.~B., {Armentrout}, B.~K., {et~al.} 2008, \apj,
  682, 94, \dodoi{10.1086/588807}

\bibitem[{{Mitsuda} {et~al.}(2007){Mitsuda}, {Bautz}, {Inoue}, {Kelley},
  {Koyama}, {Kunieda}, {Makishima}, {Ogawara}, {Petre}, {Takahashi}, {Tsunemi},
  {White}, {Anabuki}, {Angelini}, {Arnaud}, {Awaki}, {Bamba}, {Boyce}, {Brown},
  {Chan}, {Cottam}, {Dotani}, {Doty}, {Ebisawa}, {Ezoe}, {Fabian}, {Figueroa},
  {Fujimoto}, {Fukazawa}, {Furusho}, {Furuzawa}, {Gendreau}, {Griffiths},
  {Haba}, {Hamaguchi}, {Harrus}, {Hasinger}, {Hatsukade}, {Hayashida}, {Henry},
  {Hiraga}, {Holt}, {Hornschemeier}, {Hughes}, {Hwang}, {Ishida}, {Ishisaki},
  {Isobe}, {Itoh}, {Iyomoto}, {Kahn}, {Kamae}, {Katagiri}, {Kataoka},
  {Katayama}, {Kawai}, {Kilbourne}, {Kinugasa}, {Kissel}, {Kitamoto}, {Kohama},
  {Kohmura}, {Kokubun}, {Kotani}, {Kotoku}, {Kubota}, {Madejski}, {Maeda},
  {Makino}, {Markowitz}, {Matsumoto}, {Matsumoto}, {Matsuoka}, {Matsushita},
  {McCammon}, {Mihara}, {Misaki}, {Miyata}, {Mizuno}, {Mori}, {Mori}, {Morii},
  {Moseley}, {Mukai}, {Murakami}, {Murakami}, {Mushotzky}, {Nagase}, {Namiki},
  {Negoro}, {Nakazawa}, {Nousek}, {Okajima}, {Ogasaka}, {Ohashi}, {Oshima},
  {Ota}, {Ozaki}, {Ozawa}, {Parmar}, {Pence}, {Porter}, {Reeves}, {Ricker},
  {Sakurai}, {Sanders}, {Senda}, {Serlemitsos}, {Shibata}, {Soong}, {Smith},
  {Suzuki}, {Szymkowiak}, {Takahashi}, {Tamagawa}, {Tamura}, {Tamura},
  {Tanaka}, {Tashiro}, {Tawara}, {Terada}, {Terashima}, {Tomida}, {Torii},
  {Tsuboi}, {Tsujimoto}, {Tsuru}, {Turner}, {Ueda}, {Ueno}, {Ueno}, {Uno},
  {Urata}, {Watanabe}, {Yamamoto}, {Yamaoka}, {Yamasaki}, {Yamashita},
  {Yamauchi}, {Yamauchi}, {Yaqoob}, {Yonetoku}, \& {Yoshida}}]{Suzaku07}
{Mitsuda}, K., {Bautz}, M., {Inoue}, H., {et~al.} 2007, \pasj, 59, S1,
  \dodoi{10.1093/pasj/59.sp1.S1}

\bibitem[{{Murphy} \& {Yaqoob}(2009)}]{MYT09}
{Murphy}, K.~D., \& {Yaqoob}, T. 2009, \mnras, 397, 1549,
  \dodoi{10.1111/j.1365-2966.2009.15025.x}

\bibitem[{{Nandra} {et~al.}(2007){Nandra}, {O'Neill}, {George}, \&
  {Reeves}}]{Nandra07}
{Nandra}, K., {O'Neill}, P.~M., {George}, I.~M., \& {Reeves}, J.~N. 2007,
  \mnras, 382, 194, \dodoi{10.1111/j.1365-2966.2007.12331.x}

\bibitem[{{Nandra} {et~al.}(2013){Nandra}, {Barret}, {Barcons}, {Fabian}, {den
  Herder}, {Piro}, {Watson}, {Adami}, {Aird}, {Afonso}, \& et~al.}]{Athena13}
{Nandra}, K., {Barret}, D., {Barcons}, X., {et~al.} 2013, ArXiv e-prints.
\newblock \doarXiv{1306.2307}

\bibitem[{{Nasa High Energy Astrophysics Science Archive Research Center
  (Heasarc)}(2014)}]{HEASoft}
{Nasa High Energy Astrophysics Science Archive Research Center (Heasarc)}.
  2014, {HEAsoft: Unified Release of FTOOLS and XANADU}, Astrophysics Source
  Code Library.
\newblock \doeprint{1408.004}

\bibitem[{{Nelson} \& {Whittle}(1995)}]{Nelson95}
{Nelson}, C.~H., \& {Whittle}, M. 1995, \apjs, 99, 67, \dodoi{10.1086/192179}

\bibitem[{{Netzer}(2015)}]{Netzer15}
{Netzer}, H. 2015, \araa, 53, 365, \dodoi{10.1146/annurev-astro-082214-122302}

\bibitem[{{Oh} {et~al.}(2011){Oh}, {Sarzi}, {Schawinski}, \& {Yi}}]{Oh11}
{Oh}, K., {Sarzi}, M., {Schawinski}, K., \& {Yi}, S.~K. 2011, \apjs, 195, 13,
  \dodoi{10.1088/0067-0049/195/2/13}

\bibitem[{{Oh} {et~al.}(2015){Oh}, {Yi}, {Schawinski}, {Koss}, {Trakhtenbrot},
  \& {Soto}}]{Oh15}
{Oh}, K., {Yi}, S.~K., {Schawinski}, K., {et~al.} 2015, \apjs, 219, 1,
  \dodoi{10.1088/0067-0049/219/1/1}

\bibitem[{{Ricci} {et~al.}(2015){Ricci}, {Ueda}, {Koss}, {Trakhtenbrot},
  {Bauer}, \& {Gandhi}}]{Ricci15}
{Ricci}, C., {Ueda}, Y., {Koss}, M.~J., {et~al.} 2015, \apjl, 815, L13,
  \dodoi{10.1088/2041-8205/815/1/L13}

\bibitem[{{Risaliti} {et~al.}(2007){Risaliti}, {Elvis}, {Fabbiano}, {Baldi},
  {Zezas}, \& {Salvati}}]{Risaliti07}
{Risaliti}, G., {Elvis}, M., {Fabbiano}, G., {et~al.} 2007, \apjl, 659, L111,
  \dodoi{10.1086/517884}

\bibitem[{{Risaliti} {et~al.}(1999){Risaliti}, {Maiolino}, \&
  {Salvati}}]{Risaliti99}
{Risaliti}, G., {Maiolino}, R., \& {Salvati}, M. 1999, \apj, 522, 157,
  \dodoi{10.1086/307623}

\bibitem[{{Risaliti} {et~al.}(2011){Risaliti}, {Salvati}, \&
  {Marconi}}]{Risaliti11}
{Risaliti}, G., {Salvati}, M., \& {Marconi}, A. 2011, \mnras, 411, 2223,
  \dodoi{10.1111/j.1365-2966.2010.17843.x}

\bibitem[{{S{\'a}nchez-Bl{\'a}zquez} {et~al.}(2006){S{\'a}nchez-Bl{\'a}zquez},
  {Peletier}, {Jim{\'e}nez-Vicente}, {Cardiel}, {Cenarro},
  {Falc{\'o}n-Barroso}, {Gorgas}, {Selam}, \& {Vazdekis}}]{Sanchez-blazquez06}
{S{\'a}nchez-Bl{\'a}zquez}, P., {Peletier}, R.~F., {Jim{\'e}nez-Vicente}, J.,
  {et~al.} 2006, \mnras, 371, 703, \dodoi{10.1111/j.1365-2966.2006.10699.x}

\bibitem[{{Schmitt} {et~al.}(2003){Schmitt}, {Donley}, {Antonucci},
  {Hutchings}, {Kinney}, \& {Pringle}}]{Schmitt03}
{Schmitt}, H.~R., {Donley}, J.~L., {Antonucci}, R.~R.~J., {et~al.} 2003, \apj,
  597, 768, \dodoi{10.1086/381224}

\bibitem[{{Shakura} \& {Sunyaev}(1973)}]{Shak73}
{Shakura}, N.~I., \& {Sunyaev}, R.~A. 1973, \aap, 24, 337

\bibitem[{{Shen} {et~al.}(2011){Shen}, {Richards}, {Strauss}, {Hall},
  {Schneider}, {Snedden}, {Bizyaev}, {Brewington}, {Malanushenko},
  {Malanushenko}, {Oravetz}, {Pan}, \& {Simmons}}]{Shen11}
{Shen}, Y., {Richards}, G.~T., {Strauss}, M.~A., {et~al.} 2011, \apjs, 194, 45,
  \dodoi{10.1088/0067-0049/194/2/45}

\bibitem[{{Terlevich} {et~al.}(1990){Terlevich}, {Diaz}, \&
  {Terlevich}}]{Terlevich90}
{Terlevich}, E., {Diaz}, A.~I., \& {Terlevich}, R. 1990, \mnras, 242, 271,
  \dodoi{10.1093/mnras/242.3.271}

\bibitem[{{Ueda} {et~al.}(2014){Ueda}, {Akiyama}, {Hasinger}, {Miyaji}, \&
  {Watson}}]{Ueda14}
{Ueda}, Y., {Akiyama}, M., {Hasinger}, G., {Miyaji}, T., \& {Watson}, M.~G.
  2014, \apj, 786, 104, \dodoi{10.1088/0004-637X/786/2/104}

\bibitem[{{Ueda} {et~al.}(2015){Ueda}, {Hashimoto}, {Ichikawa}, {Ishino},
  {Kniazev}, {V{\"a}is{\"a}nen}, {Ricci}, {Berney}, {Gandhi}, {Koss},
  {Mushotzky}, {Terashima}, {Trakhtenbrot}, \& {Crenshaw}}]{Ueda15}
{Ueda}, Y., {Hashimoto}, Y., {Ichikawa}, K., {et~al.} 2015, \apj, 815, 1,
  \dodoi{10.1088/0004-637X/815/1/1}

\bibitem[{{Veilleux} \& {Osterbrock}(1987)}]{Veilleux87}
{Veilleux}, S., \& {Osterbrock}, D.~E. 1987, \apjs, 63, 295,
  \dodoi{10.1086/191166}

\bibitem[{{Weisskopf} {et~al.}(2000){Weisskopf}, {Tananbaum}, {Van Speybroeck},
  \& {O'Dell}}]{CXO00}
{Weisskopf}, M.~C., {Tananbaum}, H.~D., {Van Speybroeck}, L.~P., \& {O'Dell},
  S.~L. 2000, in \procspie, Vol. 4012, X-Ray Optics, Instruments, and Missions
  III, ed. J.~E. {Truemper} \& B.~{Aschenbach}, 2--16

\bibitem[{{Wu} {et~al.}(2011){Wu}, {Zhang}, {Liang}, {Zhang}, \& {Zhao}}]{wu11}
{Wu}, Y.-Z., {Zhang}, E.-P., {Liang}, Y.-C., {Zhang}, C.-M., \& {Zhao}, Y.-H.
  2011, \apj, 730, 121, \dodoi{10.1088/0004-637X/730/2/121}

\bibitem[{{Yaqoob}(2012)}]{Yaqoob12}
{Yaqoob}, T. 2012, \mnras, 423, 3360, \dodoi{10.1111/j.1365-2966.2012.21129.x}

\end{thebibliography}

\end{document}